\documentclass[
  reprint,
  amsmath,amssymb,
  aps,
  prd,
  nofootinbib,
  longbibliography
]{revtex4-2}

\usepackage{graphicx}
\usepackage{bm}
\usepackage{physics}
\usepackage{hyperref}
\usepackage{color}
\usepackage{slashed}
\usepackage{mathtools}
\usepackage{multirow}
\usepackage{float}

\hypersetup{
  colorlinks=true,
  linkcolor=blue,
  citecolor=blue,
  urlcolor=blue
}

\begin{document}

\title{Approximate Energy-Integration Method for Identifying Collisional Neutrino Flavor Instabilities}

\author{Jiabao Liu}
\email{kamaelliu@asagi.waseda.jp}
\affiliation{Department of Physics, Waseda University, Tokyo 169--8555, Japan}
\author{Hiroki Nagakura}
\affiliation{Division of Science, National Astronomical Observatory of Japan, 2-21-1 Osawa, Mitaka, Tokyo 181-8588, Japan}


\begin{abstract}
We present an approximate energy-integration method for identifying collisional neutrino flavor instabilities. Direct evaluation of the dispersion relation requires multi-dimensional integrals over neutrino phase space, making systematic searches for unstable modes in numerical models of core-collapse supernovae (CCSNe) and binary neutron star mergers (BNSMs) computationally expensive. In the literature there are some approximate schemes, but they are largely restricted to the homogeneous limit and can exhibit inaccuracies as reported in recent studies. In the current paper, we clarify the origin of the limitations in previous schemes and provide a better approximation method that robustly preserves the key physics of spectral asymmetries and collision rates. It yields a reduced dispersion relation that is inexpensive to evaluate. Comparison with exact solutions demonstrates that our new approximate method shows a good performance in computing both real frequencies and growth rates across a wide range of regimes, including isotropic and anisotropic neutrino distributions for both homogeneous and inhomogeneous modes. This provides a practical, accurate, and scalable framework for identifying collisional flavor instabilities in high-energy astrophysical simulations such as CCSNe and BNSMs.
\end{abstract}


\maketitle

\section{Introduction}

Neutrino flavor conversion in dense astrophysical environments such as core-collapse supernovae (CCSNe) and binary neutron star mergers (BNSMs) has emerged as a central problem in modern neutrino astrophysics~\cite{universe8020094,annurev:/content/journals/10.1146/annurev-nucl-102920-050505,annurev:/content/journals/10.1146/annurev-nucl-121423-100853,RevModPhys.96.025004,YAMADA2024pjab.100.015,FISCHER2024104107,raffelt2026neutrinoscorecollapsesupernovae,janka2026longtermmultidimensionalmodelscorecollapse}. In these systems, neutrinos are not only abundant but also strongly coupled through forward scattering~\cite{PhysRevD.48.1462,PANTALEONE1992128}, giving rise to collective flavor evolution phenomena that can proceed on timescales much shorter than those associated with vacuum oscillations or relevant hydrodynamics. These processes have important implications for the neutrino radiation field, and consequently for the dynamics, nucleosynthesis, and observable signals of these events. 

A variety of flavor instabilities have been identified in such environments. While early studies focused on ``slow'' modes driven by neutrino mass differences~\cite{annurev:/content/journals/10.1146/annurev.nucl.012809.104524,Duan_2009}, it was later recognized that anisotropic angular distributions can trigger ``fast'' flavor instabilities (FFI)~\cite{PhysRevD.72.045003,PhysRevD.79.105003}, whose growth rates are set by the neutrino self-interaction potential. The discovery of FFI has motivated numerous studies of their physical mechanisms and astrophysical relevance in CCSNe and BNSMs~\cite{Richers2020,PhysRevD.95.103007,PhysRevD.100.043004,PhysRevResearch.2.012046,PhysRevD.101.043016,PhysRevD.104.083025,PhysRevD.106.083005,GROHS2023138210,PhysRevD.106.043011,PhysRevD.109.043046,Mukhopadhyay_2024,PhysRevD.109.123008,PhysRevLett.134.051003,Fiorillo:2024fastI,Fiorillo:2024fastII,liu2025dynamicalequilibriafastneutrino,htq2-d1t7,Nagakura_2019,PhysRevLett.130.211401,PhysRevLett.129.261101,PhysRevD.109.023012,PhysRevD.107.103022,PhysRevLett.133.221004,PhysRevD.104.103003,PhysRevD.104.103023,PhysRevD.101.043009,PhysRevLett.126.061302}.

It has been recently recognized that collisions can induce flavor instabilities (collisional flavor instability, CFI)~\cite{PhysRevLett.130.191001}. The discovery of CFI and its resonance behavior~\cite{PhysRevD.108.083002,PhysRevD.107.123011} have stimulated extensive studies of their physical mechanisms~\cite{PhysRevD.106.103031,PhysRevD.106.103029,PhysRevD.107.083034,PhysRevD.108.083002,PhysRevD.107.123011,PhysRevD.109.063021} and astrophysical relevance in CCSNe~\cite{PhysRevD.107.083016,PhysRevD.108.123024,PhysRevD.109.023012,PhysRevD.109.103011,zaizen2024inspectingneutrinoflavorinstabilities,z3qh-nj18} and BNSMs~\cite{htq2-d1t7,PhysRevD.108.083002,nagakura2025neutrinoflavorinstabilitiesbinary}.

These flavor instabilities can be analyzed within a unified framework based on the linearization of the quantum kinetic equations (QKEs), leading to a dispersion relation (DR) that determines the spectrum of unstable modes~\cite{PhysRevD.84.053013,Airen_2018}. In general, this DR depends on both the angular and energy distributions of neutrinos, and its exact evaluation requires multi-dimensional integrals over the full neutrino phase space. While the angular structure can often be treated using moment-based closures or discrete angular bins, the energy dependence introduces an additional layer of complexity. As a result, the multi-energy dispersion relation becomes not only computationally expensive to evaluate, but also difficult to solve systematically for all relevant eigenmodes. This poses a significant challenge for large-scale surveys of unstable solutions across parameter space.

To address this difficulty for CFI stability analysis, various approximation schemes have been developed that reduce the multi-energy problem to an effective few-energy-bin description~\cite{PhysRevD.107.083034,PhysRevD.108.083002,PhysRevD.107.123011}. These approaches typically rely on energy-averaged quantities, such as effective number densities or collision rates, and have been widely used in large-scale surveys~\cite{PhysRevD.108.123024,PhysRevD.109.023012,nagakura2025neutrinoflavorinstabilitiesbinary,htq2-d1t7}. However, such reductions are generally restricted to the homogeneous ($k=0$) limit and exhibit certain inaccuracies in realistic regimes~\cite{z3qh-nj18,htq2-d1t7}.

These limitations indicate that a reliable reduction of the dispersion relation must retain the essential energy-dependent structure that controls the instability, rather than relying on simple averaging. In this work, we develop a new approximate energy-integration method that reduces the multi-energy dispersion relation to a compact form while preserving the key physics of CFI. The method expresses the energy dependence in terms of a small set of effective spectral quantities, leading to a reduced DR that is both inexpensive to evaluate and applicable beyond the $k=0$ limit. By construction, it avoids spurious singularities and mitigates the biases inherent in previous approaches. we assess the performance of the proposed method by comparing to exact solutions in a variety of setups, including isotropic and anisotropic neutrino distributions. We show that the method provides accurate estimates of both the real frequencies and growth rates of unstable modes across a broad range of models. 

This paper is organized as follows. In Sec.~\ref{sec:QKE}, we review the quantum kinetic equations and their linearization, and formulate the general dispersion relation. In Sec.~\ref{sec:Methods_AB}, we summarize existing energy-reduction approaches in the isotropic $k=0$ limit and discuss their limitations. In Sec.~\ref{sec:new_method}, we introduce the new approximate energy-integration method and extend it to anisotropic distributions and inhomogeneous modes. In Sec.~\ref{sec:Tests}, we assess the performance of the method through a series of numerical experiments, and analyze the origin of its mode-dependent accuracy. Finally, we summarize our findings and outline future directions in Sec.~\ref{sec:Conclusion}.

\section{The Quantum Kinetic Equations and Linear Stability Analysis}
\label{sec:QKE}

The neutrino flavor content in the two-flavor framework is described by the neutrino flavor density matrix
\begin{equation}
    \rho(x,P)=\begin{pmatrix}
        f_{\nu_e} & S \\
        S^* & f_{\nu_x},
    \end{pmatrix}
\end{equation}
where the star denotes complex conjugation. The diagonal components $f_{\nu_\alpha}$ represent the neutrino distribution functions in the flavor eigenstates $\alpha$, while the off-diagonal element $S\in\mathbb{C}$ encodes flavor coherence. Here $x=(x^\mu)$ is the spacetime coordinate and $P=(E,\mathbf{v})$ is the neutrino four-momentum, where neutrinos are treated as massless relativistic particles with $|\mathbf{v}|=1$ in this study. Natural units are used throughout. The barred density matrix $\bar{\rho}$ describes antineutrinos, which we treat as independent degrees of freedom\footnote{A common alternative convention treats antineutrinos as negative-energy states via $\rho(E)=-\bar{\rho}(E)$. The choice does not affect the linear stability analysis presented here.}. We adopt the Minkowski metric $\eta^{\mu\nu}=\mathrm{diag}(+1,-1,-1,-1)$.

The evolution of the flavor density matrix is governed by the quantum kinetic equations \cite{SIGL1993423,PhysRevD.94.033009,PhysRevD.99.123014,Froustey_2020}
\begin{equation}
\begin{split}
    iv\cdot\partial\rho=&[H,\rho]+iC,\\
    iv\cdot\partial\bar{\rho}=&[H,\bar{\rho}]+i\bar{C},
\end{split}
\label{eq:QKE}
\end{equation}
where $H$ is the neutrino oscillation Hamiltonian and $C$ is the collision term. In this work we neglect vacuum and matter contributions in the Hamiltonian, which do not contribute to CFI, while we focus on the neutrino self-interaction term
\begin{equation}
    H=H_{\nu\nu}(x,P)=\sqrt{2}G_{\text{F}}\,v\cdot\int dP'\left[\rho(x,P')-\bar{\rho}(x,P')\right]v',
\end{equation}
where the integration measure is defined as
\begin{equation}
    \int dP=\int_0^\infty\frac{E^2\,dE}{2\pi^2}\int\frac{d\mathbf{v}}{4\pi}.
\end{equation}
Both neutrinos and antineutrinos evolve under the similar Hamiltonian (see also~\cite{PhysRevD.106.063011}).

The full collision term is, in general, given by integrals over interaction kernels. For the purpose of linear stability analysis, we adopt a relaxation approximation,
\begin{equation}
    C(x,P)=\frac{1}{2}\{\mathrm{diag}(\Gamma_{\nu_e}(x,P),\,\Gamma_{\nu_x}(x,P)),\,\rho_{\mathrm{eq}}(x,P)-\rho(x,P)\},
\end{equation}
where $\{\cdot,\cdot\}$ denotes the anticommutator, $\Gamma_{\nu_\alpha}$ is the collision rate for flavor $\alpha$, and $\rho_{\mathrm{eq}}$ is the equilibrium density matrix toward which collisions drive the system; practically, we assume the initial flavor eigenstate with vanishing flavor coherence to be the equilibrium for the purpose of linear stability analysis. An analogous expression applies to antineutrinos.

Flavor instabilities correspond to eigenmodes of the flavor coherence $S$ that grow exponentially in time. We adopt a plane-wave ansatz,
\begin{equation}
    S(x,P)=S(k,P)e^{-ik\cdot x},\quad \bar{S}(x,P)=\bar{S}(k,P)e^{-ik\cdot x},
\end{equation}
with four-wavevector $k=(\omega,\mathbf{k})$. Linearizing Eq.~\eqref{eq:QKE} with respect to $S$ yields
\begin{equation}
\begin{split}
    [v\cdot(k-\Phi)+i\Gamma(P)]S(k,P)+\Delta f\,v\cdot a=&0,\\
    [v\cdot(k-\Phi)+i\bar{\Gamma}(P)]\bar{S}(k,P)+\Delta\bar{f}\,v\cdot a=&0,
\end{split}
\label{eq:QKE_lin}
\end{equation}
where we have defined the distribution differences
\begin{equation}
    \Delta f\equiv f_{\nu_e}-f_{\nu_x},\quad \Delta\bar{f}\equiv f_{\bar{\nu}_e}-f_{\bar{\nu}_x},
\end{equation}
the mean-field neutrino potential
\begin{equation}
    \Phi\equiv\sqrt{2}G_{\text{F}}\int dP(\Delta f-\Delta\bar{f})\,v,
\end{equation}
and the collective coherence four-vector
\begin{equation}
    a\equiv\sqrt{2}G_{\text{F}}\int dP(S-\bar{S})\,v.
    \label{eq:def_a}
\end{equation}
The collision rates appear above are the averages of those of the two neutrino flavors:
\begin{equation}
    \Gamma(P)\equiv\frac{\Gamma_{\nu_e}(P)+\Gamma_{\nu_x}(P)}{2},\quad\bar{\Gamma}(P)=\frac{\Gamma_{\bar{\nu}_e}(P)+\Gamma_{\bar{\nu}_x}(P)}{2},
\end{equation}
which are formal results of the anticommutator structure in the relaxation approximation. In the following, $\Phi$ is absorbed into a real shift of $k$, which does not affect flavor instabilities.

The formal solution of Eq.~\eqref{eq:QKE_lin} can be written as
\begin{equation}
\begin{split}
    S(k,P)=&-\frac{\Delta f(P)\,v\cdot a}{v\cdot k+i\Gamma(P)},\\
    \bar{S}(k,P)=&-\frac{\Delta\bar{f}(P)\,v\cdot a}{v\cdot k+i\bar{\Gamma}(P)}.
\end{split}
\label{eq:formal_sol_S}
\end{equation}
Substituting these expressions into Eq.~\eqref{eq:def_a}, we obtain a homogeneous equation for $a$,
\begin{equation}
\Pi^{\mu\nu}(k)\, a_\nu = 0 ,
\label{eq:Pi_eq}
\end{equation}
where the polarization tensor is given by
\begin{widetext}
\begin{equation}
\Pi^{\mu\nu}(k)
= \eta^{\mu\nu}
+ \sqrt{2}\,G_{\!F}
\int dP\,
\frac{\Delta f(P)\,v^\mu v^\nu}{\omega-\mathbf{v}\cdot\mathbf{k} + i\,\Gamma(P)}
- \sqrt{2}\,G_{\!F}
\int dP\,
\frac{\Delta\bar{f}(P)\,v^\mu v^\nu}{\omega-\mathbf{v}\cdot\mathbf{k} + i\,\bar{\Gamma}(P)}\,.
\label{eq:Pi_def}
\end{equation}
\end{widetext}

Nontrivial solutions for $a$ exist if and only if
\begin{equation}
\det \Pi^{\mu\nu}(k) = 0 ,
\label{eq:DR}
\end{equation}
which defines the dispersion relation $\omega(\mathbf{k})$. An instability is present whenever a solution satisfies $\mathrm{Im}\,\omega>0$ for some wavevector $\mathbf{k}$.

The evaluation of Eq.~\eqref{eq:DR} requires multi-dimensional integrals over energy and angle, whose convergence typically demands very fine resolution. In addition, solving the dispersion relation numerically poses a nontrivial root-finding problem. The solutions correspond to isolated zeros of $\det\Pi^{\mu\nu}(\omega,\mathbf{k})$ in the complex $\omega$ plane, which may appear as sharp, localized features that are difficult to locate and can be easily missed by standard iterative methods.

These challenges make a direct numerical approach both computationally expensive and operationally cumbersome. A well-constructed approximation method can therefore provide accurate initial estimates of the relevant modes while also serving as an efficient alternative to solving the full dispersion relation. In the following section, we introduce a robust approximate method that enables efficient evaluation of the dispersion relation while retaining the essential physics of the instability.

\section{Previous Approximations in the Isotropic $k=0$ Limit}
\label{sec:Methods_AB}

The central computational difficulty in solving the dispersion relation defined in Eq.~\eqref{eq:DR} lies in the phase-space dependence of the polarization tensor, in particular the energy-angle-dependent denominators appearing in the dispersion integrals. Accurate evaluation of these integrals during the root-search process can be computationally demanding. Moreover, locating unstable solutions of the exact dispersion relation in the complex frequency plane may be operationally challenging. These considerations motivate reduced energy-integration methods that preserve the physically relevant unstable modes while substantially simplifying the numerical problem. In this section, we review two commonly used approximations for CFI stability analysis in the isotropic $k=0$ limit. We then introduce in Sec.~\ref{sec:new_method} a new, more robust method and extend it to anisotropic angular distributions and nonzero wavevectors.

The two previous approximate schemes deal with isotropic neutrino distributions and the $k=0$ mode, which provide the simplest setting in which collisional flavor instabilities arise. In this limit, the main approximation problem is the treatment of the energy dependence in the collision rates $\Gamma(E)$ and $\bar{\Gamma}(E)$.

For isotropic backgrounds, the dispersion relation reduces to
\begin{equation}
    1+\sqrt{2}G_F\int_0^\infty\frac{E^2\,dE}{2\pi^2}
    \left[
        \frac{\Delta f(E)}{\omega+i\Gamma(E)}
        -
        \frac{\Delta \bar{f}(E)}{\omega+i\bar{\Gamma}(E)}
    \right]=0
\label{eq:DR_IP_k0}
\end{equation}
for the $\mu\nu=00$ component of Eq.~\eqref{eq:DR}, and
\begin{equation}
    -3+\sqrt{2}G_F\int_0^\infty\frac{E^2\,dE}{2\pi^2}
    \left[
        \frac{\Delta f(E)}{\omega+i\Gamma(E)}
        -
        \frac{\Delta \bar{f}(E)}{\omega+i\bar{\Gamma}(E)}
    \right]=0
\label{eq:DR_IB_k0}
\end{equation}
for the spatial diagonal components $\mu\nu=11,22,33$. The former corresponds to the isotropy-preserving (IP) mode, while the latter corresponds to the isotropy-breaking (IB) mode~\cite{PhysRevD.107.123011}.

Two closely related approximation schemes have been widely used in this setting. Both replace the energy-dependent collision rates by effective constants, thereby reducing the dispersion relation to a quadratic equation in $\omega$. The first scheme, referred to here as method~A \cite{PhysRevD.107.083034,PhysRevD.108.083002} proposed in, defines
\begin{equation}
\begin{split}
\Gamma^A_{\text{eff}}&=
\frac{\int_0^\infty E^2\,dE\,\Delta f(E)\,\Gamma(E)}
     {\int_0^\infty E^2\,dE\,\Delta f(E)},\\
\bar{\Gamma}^A_{\text{eff}}&=
\frac{\int_0^\infty E^2\,dE\,\Delta\bar f(E)\,\bar{\Gamma}(E)}
     {\int_0^\infty E^2\,dE\,\Delta\bar f(E)}.
\end{split}
\label{eq:Gamma_methodA}
\end{equation}

A second scheme, referred to here as method~B \cite{PhysRevD.107.123011}, first computes flavor-dependent effective collision rates,
\begin{equation}
\Gamma_{\text{eff},\nu_\alpha}=
\frac{\int_0^\infty E^2\,dE\,f_{\nu_\alpha}(E)\,\Gamma_{\nu_\alpha}(E)}
     {\int_0^\infty E^2\,dE\,f_{\nu_\alpha}(E)},
\end{equation}
with an analogous definition for antineutrinos, and then defines
\begin{equation}
\Gamma^B_{\text{eff}}=\frac{\Gamma_{\text{eff},\nu_e}+\Gamma_{\text{eff},\nu_x}}{2},
\qquad
\bar{\Gamma}^B_{\text{eff}}=\frac{\Gamma_{\text{eff},\bar{\nu}_e}+\Gamma_{\text{eff},\bar{\nu}_x}}{2}.
\end{equation}

Defining
\begin{equation}
\begin{split}
g&\equiv\sqrt{2}G_{\!F}\int_0^\infty\frac{E^2\,dE}{2\pi^2}\,\Delta f(E),\\
\bar g&\equiv\sqrt{2}G_{\!F}\int_0^\infty\frac{E^2\,dE}{2\pi^2}\,\Delta\bar f(E),
\end{split}
\end{equation}
the reduced IP dispersion relation in both methods can be written as
\begin{equation}
1+\left(
\frac{g}{\omega+i\Gamma_{\text{eff}}^{X}}
-
\frac{\bar g}{\omega+i\bar{\Gamma}_{\text{eff}}^{X}}
\right)=0 ,
\label{eq:AB_DR_IP}
\end{equation}
where $X\in\{\mathrm{A},\mathrm{B}\}$. The corresponding solutions are
\begin{equation}
\omega^X_\pm=-A^X-i\gamma^X\pm\sqrt{(A^X)^2-(\alpha^X)^2+i\,2G^X\alpha^X},
\label{eq:mono_AB_IP}
\end{equation}
with
\begin{widetext}
\begin{equation}
G^X\equiv\frac{g+\bar g}{2},\qquad
A^X\equiv\frac{g-\bar g}{2},\qquad
\gamma^X\equiv\frac{\Gamma_{\text{eff}}^X+\bar{\Gamma}_{\text{eff}}^X}{2},\qquad
\alpha^X\equiv\frac{\Gamma_{\text{eff}}^X-\bar{\Gamma}_{\text{eff}}^X}{2}.
\end{equation}
\end{widetext}
The reduced IB dispersion relation in both methods can be obtained by replacing $1\rightarrow-3$ in Eq.~\eqref{eq:AB_DR_IP}:
\begin{equation}
    -3+\left(
\frac{g}{\omega+i\Gamma_{\text{eff}}^{X}}
-
\frac{\bar g}{\omega+i\bar{\Gamma}_{\text{eff}}^{X}}
\right)=0 ,
\label{eq:AB_DR_IB}
\end{equation}
whose solution is given by
\begin{equation}
    \omega^X_\pm=\frac{A^X}{3}-i\gamma^X\pm\sqrt{\left(\frac{A^X}{3}\right)^2-(\alpha^X)^2-i\frac{2}{3}G^X\alpha^X}.
\label{eq:mono_AB_IB}
\end{equation}

The solutions of Eqs.~\eqref{eq:mono_AB_IP} and \eqref{eq:mono_AB_IB} admit two commonly used limiting forms,
\begin{equation}
\begin{split}
&\omega^X_\pm\approx \\
&
\begin{cases}
-A^X-i\gamma^X\pm\left(|A^X|+i\,\dfrac{G^X\alpha^X}{|A^X|}\right), & (A^X)^2\gg|G^X\alpha^X|,\\[6pt]
-A^X-i\gamma^X\pm\sqrt{i\,2G^X\alpha^X}, & (A^X)^2\ll|G^X\alpha^X|,
\end{cases}
\label{eq:mono_lim_AB_IP}
\end{split}
\end{equation}
for the IP branch, and 
\begin{equation}
\begin{split}
&\omega^X_\pm\approx \\
&
\begin{cases}
\frac{A^X}{3}-i\gamma^X\pm\left(\frac{|A^X|}{3}-i\,\dfrac{G^X\alpha^X}{|A^X|}\right), & (A^X)^2\gg|G^X\alpha^X|,\\[6pt]
\frac{A^X}{3}-i\gamma^X\pm\sqrt{-i\,\frac{2}{3}G^X\alpha^X}, & (A^X)^2\ll|G^X\alpha^X|,
\end{cases}
\label{eq:mono_lim_AB_IB}
\end{split}
\end{equation}
for the IB branch. In the above expansions, the upper and lower rows correspond to the non-resonance and resonance limits, respectively. In the non-resonance regime, the solution pair typically separates into a near mode, defined by $\mathrm{Re}\,\omega \approx 0$, and a far mode, whose real parts of the eigenfrequencies are $|\mathrm{Re}\,\omega| \sim 2|A|$. Hereafter, we adopt this terminology, near and far modes, throughout.

Both methods reproduce certain qualitative features of the exact dispersion relation and can provide order-of-magnitude estimates of growth rates in favorable cases, unless the solutions hit singularities (see below for more details). Their simplicity also makes them attractive for exploratory applications. However, both methods have important limitations, whose details are discussed below.\footnote{Recently, Froustey et al.~\cite{htq2-d1t7} assessed the accuracy of methods~A and~B using BNSM data and proposed a hybrid approach combining elements of the two. Wang et al.~\cite{z3qh-nj18} also showed, based on CCSN data, that method~B can yield inaccurate results.}

Method~A can be formally derived by expanding and re-summing the denominator:
\begin{widetext}
\begin{equation}
\begin{split}
\int_0^\infty E^2\,dE\,
\frac{\Delta f(E)}{\omega\left[1+i\Gamma(E)/\omega\right]}
&\approx
\int_0^\infty E^2\,dE\,
\frac{\Delta f(E)}{\omega}
\left(1-i\frac{\Gamma(E)}{\omega}\right)\\
&=
\int_0^\infty E^2\,dE\,
\frac{\Delta f(E)}{\omega}
\left(1-i\frac{\Gamma^A_{\text{eff}}}{\omega}\right)\\
&\approx
\frac{\int_0^\infty E^2\,dE\,\Delta f(E)}
{\omega+i\Gamma^A_{\text{eff}}}.
\end{split}
\end{equation}
\end{widetext}
However, this approximation can break down when $\Delta f(E)$ exhibits multiple zero-crossings in energy. In such cases, cancellations in the denominator of Eq.~\eqref{eq:Gamma_methodA} can render $\Gamma^A_{\text{eff}}$ ill-defined or even negative, and even when finite, it may become much larger than the microscopic rates entering the original integral, thereby invalidating the expansion itself. This pathology underlies parts of the inaccurate CFI growth rates reported in~\cite{htq2-d1t7}. Method~B, by contrast, avoids divergent effective collision rates by construction, but it is essentially empirical and does not follow from a controlled reorganization of the dispersion integrals; as a result, it can overestimate CFI growth rates~\cite{z3qh-nj18} and, as we show in Sec.~\ref{sec:Tests}, may also yield quantitatively inaccurate results.

These limitations motivate the development of a more robust approximation scheme that retains the essential phase-space information while remaining as simple as the methods reviewed above for practical use.

\section{A New Approximate Method}
\label{sec:new_method}
In this section, we introduce a new energy-reduction scheme that overcomes the limitations of the previous approximate methods. Consistent with the naming convention of methods~A and B, we refer to the new one as method~C. We begin with the simplest setting, namely isotropic angular distributions in the $k=0$ limit, in Sec.~\ref{subsec:MethodC_Isok0}. We then extend the construction to anisotropic angular distributions and inhomogeneous modes in Sec.~\ref{subsec:MethodC_extension}.

\subsection{Isotropic Distributions and $k=0$ Mode}
\label{subsec:MethodC_Isok0}
For any real-valued function $X$, we define
\begin{equation}
    [X]_{+} \equiv \max\,\!\bigl(X,0\bigr),
    \qquad
    [X]_{-} \equiv \max\,\!\bigl(-X,0\bigr),
\end{equation}
so that
\begin{equation}
X=[X]_{+}-[X]_{-},
\end{equation}
with both parts $[X]_\pm$ nonnegative.

\begin{figure}[h]
\centering
\includegraphics[width=\linewidth]{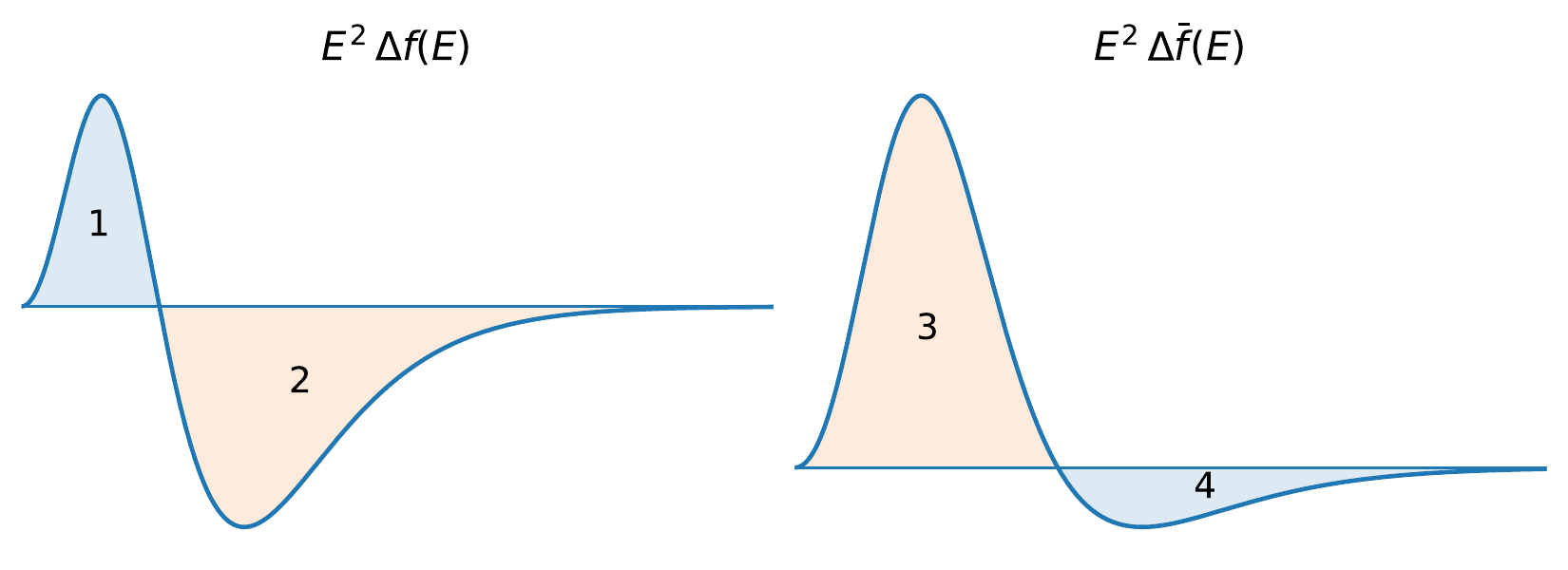}
\caption{Schematic decomposition of $E^{2}\Delta f(E)$ and $E^{2}\Delta\bar f(E)$ into four signed contributions (labeled $1$--$4$). The effective densities entering methods~A/B and the present method are constructed from different combinations of these areas.}
\label{fig:cross_scheme}
\end{figure}

Applying this decomposition to the neutrino and antineutrino flavor-lepton-number distributions, we can write
\begin{equation}
\begin{split}
\Delta f(E)&=[\Delta f(E)]_{+}-[\Delta f(E)]_{-},\\
\Delta\bar f(E)&=[\Delta\bar f(E)]_{+}-[\Delta\bar f(E)]_{-}.
\end{split}
\end{equation}
Because neutrino and antineutrino contributions enter the $k=0$ dispersion relation with opposite signs, it is natural to group
\begin{equation}
[\Delta f(E)]_+ \quad \text{and} \quad [\Delta\bar f(E)]_-
\end{equation}
into an effective positive sector, and
\begin{equation}
[\Delta f(E)]_- \quad \text{and} \quad [\Delta\bar f(E)]_+
\end{equation}
into an effective negative sector. This organization avoids cancellations between opposite-sign contributions within the same effective sector. 

We expand the denominators to first order and obtain
\begin{widetext}
\begin{equation}
\begin{split}
&\sqrt{2}G_{\!F}\int_0^\infty \frac{E^2\,dE}{2\pi^2}
\left[
\frac{\Delta f(E)}{\omega+i\Gamma(E)}
-
\frac{\Delta\bar f(E)}{\omega+i\bar\Gamma(E)}
\right]\\
\approx\;&
\frac{\sqrt{2}G_{\!F}}{\omega}
\int_0^\infty \frac{E^2\,dE}{2\pi^2}
\Bigg\{
\Big[
[\Delta f(E)]_{+}\Big(1-i\frac{\Gamma(E)}{\omega}\Big)
+
[\Delta\bar f(E)]_{-}\Big(1-i\frac{\bar\Gamma(E)}{\omega}\Big)
\Big]\\
&\hspace{5.3cm}-
\Big[
[\Delta f(E)]_{-}\Big(1-i\frac{\Gamma(E)}{\omega}\Big)
+
[\Delta\bar f(E)]_{+}\Big(1-i\frac{\bar\Gamma(E)}{\omega}\Big)
\Big]
\Bigg\}.
\end{split}
\end{equation}
\end{widetext}

It is then convenient to define the sector-wise moments
\begin{equation}
\begin{split}
\mathcal N_{+}&\equiv
\int_0^\infty \frac{E^2\,dE}{2\pi^2}
\Big([\Delta f(E)]_{+}+[\Delta\bar f(E)]_{-}\Big),\\
\mathcal N_{-}&\equiv
\int_0^\infty \frac{E^2\,dE}{2\pi^2}
\Big([\Delta f(E)]_{-}+[\Delta\bar f(E)]_{+}\Big),
\end{split}
\end{equation}
and the corresponding collision-weighted moments
\begin{equation}
\begin{split}
\mathcal G_{+}&\equiv
\int_0^\infty \frac{E^2\,dE}{2\pi^2}
\Big([\Delta f(E)]_{+}\Gamma(E)+[\Delta\bar f(E)]_{-}\bar\Gamma(E)\Big),\\
\mathcal G_{-}&\equiv
\int_0^\infty \frac{E^2\,dE}{2\pi^2}
\Big([\Delta f(E)]_{-}\Gamma(E)+[\Delta\bar f(E)]_{+}\bar\Gamma(E)\Big).
\end{split}
\end{equation}
By construction, all four quantities are nonnegative. We therefore define sector-wise effective collision rates,
\begin{equation}
\Gamma_{+}\equiv\frac{\mathcal G_{+}}{\mathcal N_{+}},
\qquad
\Gamma_{-}\equiv\frac{\mathcal G_{-}}{\mathcal N_{-}},
\end{equation}
which remain nonnegative and well defined even in the presence of energy crossings.

With these definitions, the dispersion integral is approximated by the reduced rational form
\begin{widetext}
\begin{equation}
\int_0^\infty \frac{E^2\,dE}{2\pi^2}
\left[
\frac{\Delta f(E)}{\omega+i\Gamma(E)}
-
\frac{\Delta\bar f(E)}{\omega+i\bar\Gamma(E)}
\right]
\approx
\frac{\mathcal N_{+}}{\omega+i\Gamma_{+}}
-
\frac{\mathcal N_{-}}{\omega+i\Gamma_{-}}.
\label{eq:methodC_rational}
\end{equation}
\end{widetext}
For the IP mode, Eq.~\eqref{eq:DR_IP_k0} then becomes
\begin{equation}
1+
\left(
\frac{g_{+}}{\omega+i\Gamma_{+}}
-
\frac{g_{-}}{\omega+i\Gamma_{-}}
\right)=0,
\label{eq:DR_IP_C}
\end{equation}
where
\begin{equation}
g_{\pm}\equiv\sqrt{2}G_F\mathcal N_{\pm}.
\end{equation}
An analogous expression holds for the IB mode upon replacing $1\to -3$:
\begin{equation}
    -3+\left(
\frac{g_{+}}{\omega+i\Gamma_{+}}
-
\frac{g_{-}}{\omega+i\Gamma_{-}}
\right)=0.
\label{eq:DR_IB_C}
\end{equation}
These reduced dispersion relations have the same algebraic form as those obtained in methods~A and~B, but with differently defined effective densities and collision rates.

The reduced IP and IB branches in method~C again admit closed-form solutions,
\begin{equation}
\omega_\pm=
-A^{C}-i\gamma^{C}
\pm
\sqrt{(A^{C})^{2}-(\alpha^{C})^{2}+i\,2G^{C}\alpha^{C}},
\label{eq:mono_C_IP}
\end{equation}
for the IP mode, and
\begin{equation}
\omega_\pm=
\frac{A^{C}}{3}-i\gamma^{C}
\pm
\sqrt{\left(\frac{A^{C}}{3}\right)^{2}-(\alpha^{C})^{2}-i\,\frac{2}{3}G^{C}\alpha^{C}},
\label{eq:mono_C_IB}
\end{equation}
for the IB mode, with the four effective parameters defined as
\begin{widetext}
\begin{equation}
G^{C}\equiv \frac{g_{+}+g_{-}}{2},\qquad
A^{C}\equiv \frac{g_{+}-g_{-}}{2},\qquad
\gamma^{C}\equiv \frac{\Gamma_{+}+\Gamma_{-}}{2},\qquad
\alpha^{C}\equiv \frac{\Gamma_{+}-\Gamma_{-}}{2}.
\label{eq:def_C}
\end{equation}
\end{widetext}
These solutions have the same algebraic form as Eqs.~\eqref{eq:mono_AB_IP} and \eqref{eq:mono_AB_IB}, but the effective densities and collision rates are defined differently and therefore encode a different approximation.

They also admit the corresponding limiting forms
\begin{equation}
\begin{split}
&\omega^C_\pm \approx \\
&
\begin{cases}
-A^C-i\gamma^C \pm \left(|A^C|+ i\,\dfrac{G^C\alpha^C}{|A^C|}\right), & (A^C)^2 \gg |G^C\alpha^C|,\\[6pt]
-A^C-i\gamma^C \pm \sqrt{i\,2G^C\alpha^C}, & (A^C)^2 \ll |G^C\alpha^C| ,
\end{cases}
\end{split}
\end{equation}
for the IP branch and 
\begin{equation}
\begin{split}
&\omega^C_\pm \approx \\
&
\begin{cases}
\frac{A^C}{3}-i\gamma^C \pm \left(\frac{|A^C|}{3}- i\,\dfrac{G^C\alpha^C}{|A^C|}\right), & (A^C)^2 \gg |G^C\alpha^C|,\\[6pt]
\frac{A^C}{3}-i\gamma^C \pm \sqrt{-i\,\frac{2}{3}G^C\alpha^C}, & (A^C)^2 \ll |G^C\alpha^C| ,
\end{cases}
\end{split}
\end{equation}
for the IB branch, where the upper and lower rows correspond to the non-resonance and resonance limits, respectively.

This construction overcomes the main shortcomings of methods~A and~B. Unlike method~A, the effective collision rates $\Gamma_\pm$ cannot become negative or ill defined through cancellations in signed integrals. Unlike method~B, the approximation follows directly from an explicit reorganization of the dispersion integral rather than an empirical averaging prescription. It therefore retains the practical simplicity of earlier approaches while providing a more robust treatment of multi-energy spectra with zero-crossings.

Although $A^C$ is defined in the same way as in methods~A and~B, the definition of $G^C$ differs. As a result, the three schemes often give similar diagnostics for proximity to the resonance condition while predicting different growth rates. This difference is illustrated schematically in Fig.~\ref{fig:cross_scheme}. In methods~A and~B, the effective densities correspond to signed combinations of the four areas, while in the present construction the positive and negative sectors are grouped differently, leading to a different definition of $G$ while preserving the same signed difference that enters $A$.

\subsection{Anisotropic angular distributions and Inhomogeneous Modes}
\label{subsec:MethodC_extension}

We now extend our approximate method to treat anisotropic distributions and inhomogeneous modes. In this work, we consider only axially symmetric distributions, while it is straightforward to extend the non-axisymmetric case. We then outline the generalization to nonzero wave number.

We choose the symmetry axis as the $z$ direction and assume that the background distributions depend on the neutrino velocity only through $v_z=\cos\theta$. We further align the wavevector with the symmetry axis, $\mathbf k=k_z\hat{\mathbf z}$, and suppress the subscript $z$, writing simply $v$ and $k$. Under these assumptions, the nonvanishing components of the polarization tensor are $\Pi^{00}$, $\Pi^{03}=\Pi^{30}$, $\Pi^{33}$, and $\Pi^{11}=\Pi^{22}$, given by
\begin{widetext}
\begin{align}
\Pi^{00}(\omega,k) &= 1
+ \sqrt{2}\,G_{\!F}
\int_0^\infty\int_{-1}^1\frac{E^2\,dE}{2\pi^2}\,\frac{dv}{2}
\left[
\frac{\Delta f(E,v)}{\omega-kv+i\Gamma(E,v)}
- \frac{\Delta\bar f(E,v)}{\omega-kv+i\bar{\Gamma}(E,v)}
\right], \label{eq:Pi00} \\[4pt]
\Pi^{03}(\omega,k) &=
\sqrt{2}\,G_{\!F}
\int_0^\infty\int_{-1}^1\frac{E^2\,dE}{2\pi^2}\,\frac{dv}{2}\,v
\left[
\frac{\Delta f(E,v)}{\omega-kv+i\Gamma(E,v)}
- \frac{\Delta\bar f(E,v)}{\omega-kv+i\bar{\Gamma}(E,v)}
\right], \label{eq:Pi03} \\[4pt]
\Pi^{33}(\omega,k) &= -1
+ \sqrt{2}\,G_{\!F}
\int_0^\infty\int_{-1}^1\frac{E^2\,dE}{2\pi^2}\,\frac{dv}{2}\,v^2
\left[
\frac{\Delta f(E,v)}{\omega-kv+i\Gamma(E,v)}
- \frac{\Delta\bar f(E,v)}{\omega-kv+i\bar{\Gamma}(E,v)}
\right], \label{eq:Pi33} \\[4pt]
\Pi^{11}(\omega,k) &= \Pi^{22}(\omega,k)
= -1
+ \sqrt{2}\,G_{\!F}
\int_0^\infty\int_{-1}^1\frac{E^2\,dE}{2\pi^2}\,\frac{dv}{2}\,\frac{1-v^2}{2}
\left[
\frac{\Delta f(E,v)}{\omega-kv+i\Gamma(E,v)}
- \frac{\Delta\bar f(E,v)}{\omega-kv+i\bar{\Gamma}(E,v)}
\right]. \label{eq:Pi11}
\end{align}
\end{widetext}

The dispersion relation factorizes as
\begin{equation}
\bigl(\Pi^{00}\Pi^{33}-(\Pi^{03})^2\bigr)(\Pi^{11})^2=0,
\label{eq:TLTr_Pi}
\end{equation}
corresponding to a temporal-longitudinal (TL) sector satisfying $\bigl(\Pi^{00}\Pi^{33}-(\Pi^{03})^2\bigr)=0$ and a doubly degenerate transverse (Tr) sector obeying $\Pi^{11}=0$. 

At $k=0$, the angular dependence closes exactly on the first three Legendre moments. Writing
\begin{equation}
f_{\nu_\alpha}(E,v)=\sum_{\ell=0}^{\infty} f_{\nu_\alpha,\ell}(E)\,P_\ell(v),
\end{equation}
with
\begin{equation}
f_{\nu_\alpha,\ell}(E)=\frac{2\ell+1}{2}\int_{-1}^{1}dv\,f_{\nu_\alpha}(E,v)P_\ell(v),
\end{equation}
and similarly for $\Delta f(E,v)$ and antineutrinos, one finds that only the $\ell=0,1,2$ moments enter the polarization tensor. The resulting components are
\begin{widetext}
\begin{align}
\Pi^{00}(\omega,0)=&\,1+\sqrt{2}G_{\!F}\int_0^\infty\frac{E^2dE}{2\pi^2}
\left[
\frac{\Delta f_0(E)}{\omega+i\Gamma(E)}
-\frac{\Delta \bar{f}_0(E)}{\omega+i\bar{\Gamma}(E)}
\right],\label{eq:Axi_Pi00}\\
\Pi^{03}(\omega,0)=&\,\sqrt{2}G_{\!F}\int_0^\infty\frac{E^2dE}{2\pi^2}
\left[
\frac{\frac13\,\Delta f_1(E)}{\omega+i\Gamma(E)}
-\frac{\frac13\,\Delta \bar{f}_1(E)}{\omega+i\bar{\Gamma}(E)}
\right],\label{eq:Axi_Pi03}\\
\Pi^{33}(\omega,0)=&\,-1+\sqrt{2}G_{\!F}\int_0^\infty\frac{E^2dE}{2\pi^2}
\left[
\frac{\frac13\,\Delta f_0(E)+\frac{2}{15}\,\Delta f_2(E)}{\omega+i\Gamma(E)}
-\frac{\frac13\,\Delta \bar{f}_0(E)+\frac{2}{15}\,\Delta \bar{f}_2(E)}{\omega+i\bar{\Gamma}(E)}
\right],\label{eq:Axi_Pi33}\\
\Pi^{11}(\omega,0)=&\,-1+\sqrt{2}G_{\!F}\int_0^\infty\frac{E^2dE}{2\pi^2}
\left[
\frac{\frac13\,\Delta f_0(E)-\frac{1}{15}\,\Delta f_2(E)}{\omega+i\Gamma(E)}
-\frac{\frac13\,\Delta \bar{f}_0(E)-\frac{1}{15}\,\Delta \bar{f}_2(E)}{\omega+i\bar{\Gamma}(E)}
\right].\label{eq:Axi_Pi11}
\end{align}
\end{widetext}

The Tr dispersion relation, $\Pi^{11}=0$, is formally identical to the isotropic $k=0$ case and can therefore be treated directly by the same reduced method. If $\Pi^{03}=0$, the TL sector also decouples and the equations $\Pi^{00}=0$ and $\Pi^{33}=0$ can each be approximated in the same manner. More generally, when $\Pi^{03}\neq0$, the TL sector remains coupled through $\Pi^{00}\Pi^{33}-(\Pi^{03})^2=0$.

The reduced construction can be formulated
by applying the same positive/negative sector decomposition to the effective energy weights appearing in each polarization component. Schematically, one writes
\begin{equation}
\Pi^{\mu\nu}\approx c^{\mu\nu}
+\frac{\mathcal N^{\mu\nu}_{+}}{\omega+i\Gamma^{\mu\nu}_{+}}
-\frac{\mathcal N^{\mu\nu}_{-}}{\omega+i\Gamma^{\mu\nu}_{-}},
\label{eq:Approx_Pi_Axi}
\end{equation}
with $c^{00}=1$, $c^{03}=0$, and $c^{11}=c^{33}=-1$. The resulting TL reduced dispersion relation is, in general, a higher-order polynomial in $\omega$, whose degree depends on the degeneracy structure of the effective collision rates. It can then be solved straightforwardly with a numerical polynomial root finder, or analytically when the polynomial degree does not exceed four.

The extension to nonzero wave number follows naturally within the same framework, and we provide a possible scheme in this paper. For a positive distribution function $f(E,v)$, consider the integral
\begin{equation}
\int_{-1}^{1}\frac{v^n\,dv}{2}\int_0^\infty\frac{E^2\,dE}{2\pi^2}
\frac{f(E,v)}{\omega-kv+i\Gamma(E,v)} .
\end{equation}
Expanding the denominator to the first order and re-summing the leading terms, we obtain 
\begin{equation}
\begin{split}
&\int_{-1}^{1}\frac{v^n\,dv}{2}\int_0^\infty\frac{E^2\,dE}{2\pi^2}
\frac{f(E,v)}{\omega-kv+i\Gamma(E,v)} \\
\approx\;&
\int_{-1}^{1}\frac{v^n\,dv}{2}\int_0^\infty\frac{E^2\,dE}{2\pi^2}
\frac{f(E,v)}{\omega-kv+i\Gamma_{\mathrm{eff}}(v)},
\end{split}
\end{equation}
where the angle-dependent effective collision rate is defined by
\begin{equation}
\Gamma_{\mathrm{eff}}(v)\equiv
\frac{\int_0^\infty E^2\,dE\,f(E,v)\,\Gamma(E,v)}
     {\int_0^\infty E^2\,dE\,f(E,v)}.
\label{eq:effGm_knon0}
\end{equation}
The reduction of general signed neutrino and antineutrino contributions then proceeds in close analogy with the $k=0$ case: one separates each effective weight into positive and negative sectors, evaluates the corresponding effective collision rates, and combines the reduced contributions into an approximate polarization tensor. Once the collision rates have been reduced to be energy independent, the energy integration collapses, leaving only the angular integration. If the effective collision rate $\Gamma_{\text{eff}}$ is isotropic, one may truncate the angular distribution $f(v)$ at finite order and perform the remaining angular integrals analytically using
\begin{widetext}
\begin{equation}
\int_{-1}^1\frac{dv}{2}\frac{v^n}{\omega'-kv} = \frac{\omega'^n}{2k^{\,n+1}} \ln\!\left(\frac{\omega'+k}{\omega'-k}\right)
- \frac{1}{k} \sum_{\substack{j=0 \\ j \equiv n-1 \pmod{2}}}^{n-1} \frac{(\omega'/k)^j}{n-j},
\qquad k \neq 0,
\end{equation}
\end{widetext}
where the principal branch of the logarithm is understood and the summation symbol denotes a summation from $j=0$ to $j=n-1$ such that $j\equiv n-1\pmod2$. Alternatively, one may discretize the angular integral directly on a finite angular mesh. It would be desirable to further compress the angular dependence by introducing a small set of effective quantities of the form $-kv_{\text{eff}}+i\Gamma_{\text{eff}}$, so that the energy-reduced integral could be approximated by a finite sum over only a few terms. Since the present paper is concerned primarily with the energy reduction relevant to CFIs, we do not pursue a dedicated angular-reduction scheme here and leave that construction to subsequent work.

Although we have presented the construction for axially symmetric backgrounds, the same energy-integration strategy is not restricted to axisymmetry and can be generalized to broader classes of angular distributions without conceptual difficulty.

\subsection{Mathematical assessment}
\label{subsec:mathass}

Our new approximate method exhibits different accuracies for different modes. Before presenting the numerical experiments, we provide a mathematical argument to explain these differences in accuracy.

\begin{figure}
    \centering
    \vspace{1em}
    \includegraphics[width=0.8\linewidth]{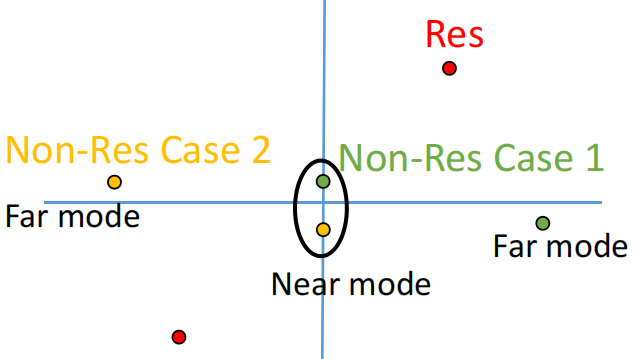}
    \caption{Schematic locations of CFI solutions in the complex $\omega$-plane. In the non-resonance regime, the solutions separate into a near mode with $\mathrm{Re}\,\omega \approx 0$ and a far mode with $|\mathrm{Re}\,\omega| \sim 2|A|$. In the resonance regime, the pair of solutions is approximately symmetric about the origin with nonzero real and imaginary parts. The figure illustrates the IP branch; the IB branch behaves analogously.}
    \label{fig:FarNearRes}
\end{figure}

A schematic illustration of the locations of CFI solutions in the complex $\omega$-plane is shown in Fig.~\ref{fig:FarNearRes}. The figure qualitatively depicts the solution pairs for the IP branch in both the non-resonance and resonance regimes; the IB branch exhibits analogous behavior and is not shown separately. In the resonance regime, the two solutions are approximately symmetric about the origin, with both real and imaginary parts nonzero. In contrast, in the non-resonance regime, one of the solutions has a vanishing real part, corresponding to the near mode defined above. In the non-resonance regime, either the near mode or the far mode can be unstable, corresponding to case 1 and case 2 illustrated in the figure, respectively.

The location of the solutions in the complex $\omega$-plane provides direct insight into the accuracy of the approximation method. In particular, modes located close to the origin (near modes) are more sensitive to the energy dependence of the collision rates, as the expansion underlying the reduced scheme involves ratios of the form $\Gamma(E)/(\omega-kv)$ and $\Gamma_{\mathrm{eff}}/(\omega-kv)$. When the relevant denominator becomes small, higher-order corrections in this expansion become non-negligible, leading to larger deviations from the exact solutions. By contrast, modes with larger characteristic frequency scale, including the far mode in the non-resonance regime and both modes in the resonance regime when sufficiently separated from the origin, are less affected by such corrections and are therefore reproduced with higher accuracy. This geometric interpretation explains the overall trends in the numerical experiments (Sec.~\ref{sec:Tests}), where the largest discrepancies are consistently associated with near modes.

A useful empirical criterion for the accuracy of the approximation is that the growth rate satisfies $|\omega|\gtrsim\Gamma_{\mathrm{eff}}$, or, more generally for $k\neq0$, that the characteristic scale of the denominator satisfies $|\omega - kv| \gtrsim \Gamma_{\mathrm{eff}}$ over the dominant angular support of the mode. When this condition holds, the leading-order expansion remains well controlled, whereas larger deviations arise when the mode approaches the origin or the resonance surface $\omega \simeq kv$. Despite these limitations, we present in Sec.~\ref{sec:Tests} that the approximation remains quantitatively reliable for practical purposes, as it correctly captures the existence, branch structure, and characteristic scales of unstable modes across a wide range of models.

\begin{table*}[t]
    \centering
    \caption{
    Parameters for the isotropic $k=0$ models discussed in Sec.~\ref{subsec:Iso_k0}. The normalization factors $g_{\nu_e}$ and $g_{\bar\nu_e}$ are fixed, while $g_{\nu_x}$ is varied to control energy crossings. In Model~II, the parameter $c_{\bar\nu_e}\simeq0.938$ is chosen such that $A^X=0$, corresponding to a resonance configuration.
    }
    \label{tab:param_iso}
    \begin{tabular}{c|cccccccc}
        \hline\hline
        Model
        & $T_{\nu_e}$ [MeV]
        & $T_{\bar\nu_e}$ [MeV]
        & $T_{\nu_x}$ [MeV]
        & $\mu_{\nu_e}$ [MeV]
        & $\mu_{\bar\nu_e}$ [MeV]
        & $\mu_{\nu_x}$ [MeV]
        & $g_{\nu_e}$
        & $g_{\bar\nu_e}$
        \\
        \hline
        I   & 3.4 & 4.5 & 5.6 & 3 & 0 & 0 & 1 & 1 \\
        II  & 3.4 & 4.5 & 5.6 & 3 & 0 & 0 & 1 & $c_{\bar\nu_e}$ \\
        \hline\hline
    \end{tabular}
\end{table*}

\section{Numerical Experiments}
\label{sec:Tests}
We test the performance and limitations of the approximate energy-integration method through controlled numerical experiments. We begin with isotropic $k=0$ collisional flavor instabilities (Sec.~\ref{subsec:Iso_k0}), comparing method~C with methods~A and~B and with exact solutions. We then consider anisotropic backgrounds at $k=0$ (Sec.~\ref{subsec:Axi_k0}), followed by extensions to $k\neq0$ for both isotropic and anisotropic distributions (Sec.~\ref{subsec:knon0}). Our goal is to assess both the quantitative accuracy of the reduced treatment and the regimes in which it provides a reliable approximation to the exact dispersion relation.

\subsection{Isotropic Distributions at $k=0$}
\label{subsec:Iso_k0}

We begin with isotropic, energy-dependent neutrino distributions and collision rates, focusing on homogeneous modes ($k=0$). This setup provides the simplest case for assessing the accuracy of the reduced energy-integration schemes.

In the present study, all neutrino species are assumed to follow scaled Fermi--Dirac energy spectra,
\begin{equation}
    f_{\nu_\alpha}(E)=g_{\nu_\alpha}\, f_{\mathrm{FD}}(E;T_\alpha,\mu_\alpha),
    \label{eq:dist_iso}
\end{equation}
where $f_{\mathrm{FD}}(E;T,\mu)$ denotes the Fermi--Dirac distribution characterized by temperature $T$ and chemical potential $\mu$, and $g_{\nu_\alpha}$ is a dimensionless normalization factor controlling the relative abundance of each flavor. 

The collision rates are taken to be energy dependent,
\begin{equation}
    \Gamma_{\nu_\alpha}(E)=\Gamma_{\nu_\alpha,0}\left(\frac{E}{10~\mathrm{MeV}}\right)^2,
    \label{eq:Gamma_E}
\end{equation}
which captures the leading energy scaling expected for neutrino-matter interactions in dense astrophysical environments. Throughout this paper we adopt
\begin{equation}
\begin{split}
\Gamma_{\nu_e,0}=1.8\times10^{-5}\,\mathrm{cm}^{-1},\\
\Gamma_{\bar\nu_e,0}=0.8\times10^{-5}\,\mathrm{cm}^{-1},\\
\Gamma_{\nu_x,0}=0.2\times10^{-5}\,\mathrm{cm}^{-1},
\end{split}
\label{eq:Gamma_params}
\end{equation}
and assume identical energy distributions and collision rate functions for $\nu_x$ and $\bar\nu_x$; i.e., we take $\Gamma_{\nu_x}(E)=\Gamma_{\bar{\nu}_x}(E)$ and $g_{\nu_x} = g_{\bar{\nu}_x}$. For all calculations presented in this work, the energy interval $[10^{-4}\,\mathrm{MeV},\,100\,\mathrm{MeV}]$ is discretized using 500 logarithmically spaced grid points. This resolution is verified to provide converged multi-energy solutions for the quantities of interest.

To vary the spectral topology in a controlled manner, we change $g_{\nu_x}$, while other parameters such as temperature, chemical potential, and $g_{\nu_e}$ and $g_{\bar\nu_e}$ are fixed in each model (Model~I and~II; see below for more details). This continuously modifies the energy spectra of $\Delta f(E)=f_{\nu_e}(E)-f_{\nu_x}(E)$ and $\Delta\bar f(E)=f_{\bar\nu_e}(E)-f_{\bar\nu_x}(E)$, thereby controlling the number and location of energy crossings. We consider two representative models with fixed parameters, summarized in Table~\ref{tab:param_iso}. Model~I corresponds to a non-resonance configuration, while Model~II is tuned to the resonance condition $A^X=0$. In both cases, we compare dispersion-relation solutions obtained using methods~A, B, and~C to multi-energy exact solutions for the IP branch. The IB branch exhibits analogous behavior and is therefore not discussed separately.

\begin{figure*}
    \centering
    \includegraphics[width=\linewidth]{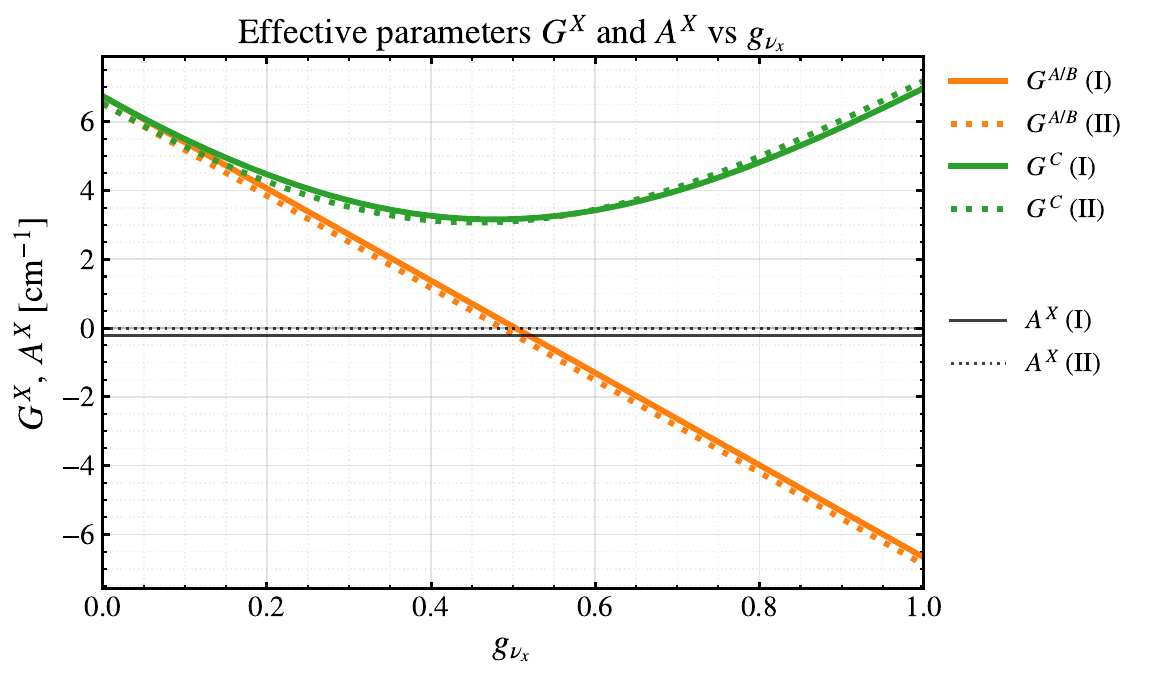}
    \caption{
    Effective coefficients $G^X$ and $A^X$ as functions of $g_{\nu_x}$ for the isotropic $k=0$ models. Model~I (solid) corresponds to a non-resonance configuration, while Model~II (dotted) is tuned to $A^X=0$.
    }
    \label{fig:GA}
\end{figure*}

We first examine the effective coefficients $G^X$ and $A^X$ as functions of $g_{\nu_x}$, shown in Fig.~\ref{fig:GA}. All methods yield identical values of $A^X$, which remain constant with respect to $g_{\nu_x}$. Model~I corresponds to $A^X\simeq -0.2\,\mathrm{cm}^{-1}$, while Model~II satisfies $A^X=0$. 

Method~C guarantees $G^C\ge0$ by construction, whereas in methods~A and~B the corresponding coefficient $G^X$ could change sign as $g_{\nu_x}$ varies. Such sign changes can interchange the growth rates of the two solutions $\omega_\pm$, leading to incorrect identification of the unstable mode in methods~A and~B (see below).

\begin{figure*}[]
    \centering
    \includegraphics[width=0.48\linewidth]{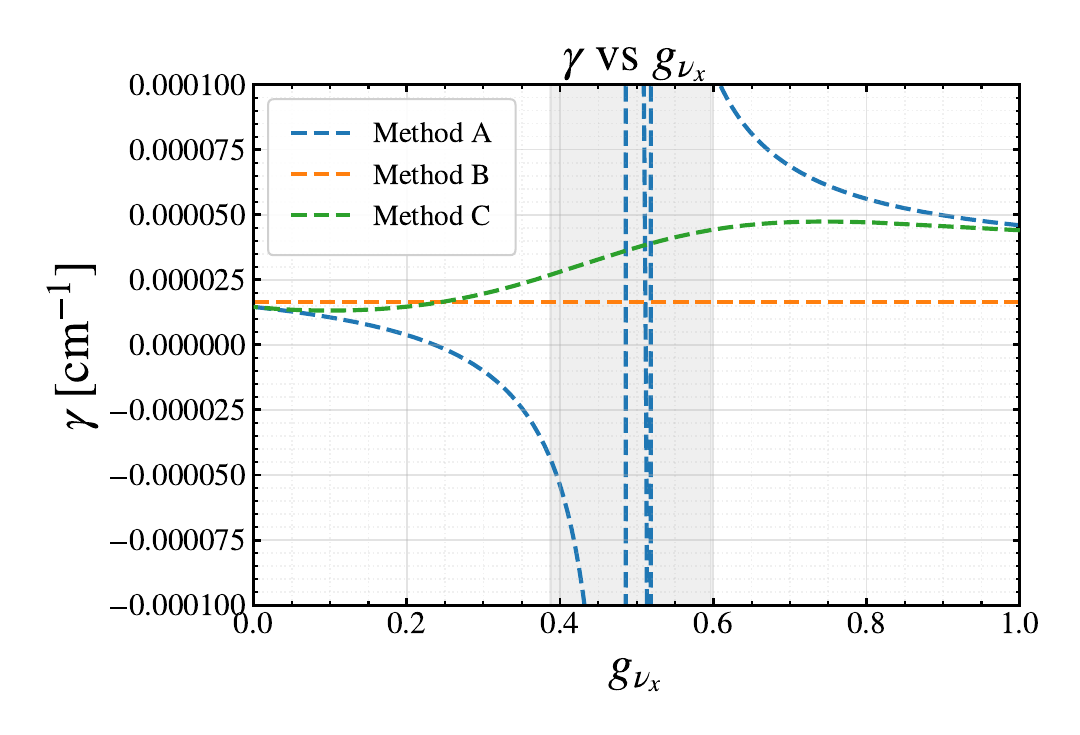}\hfill
    \includegraphics[width=0.48\linewidth]{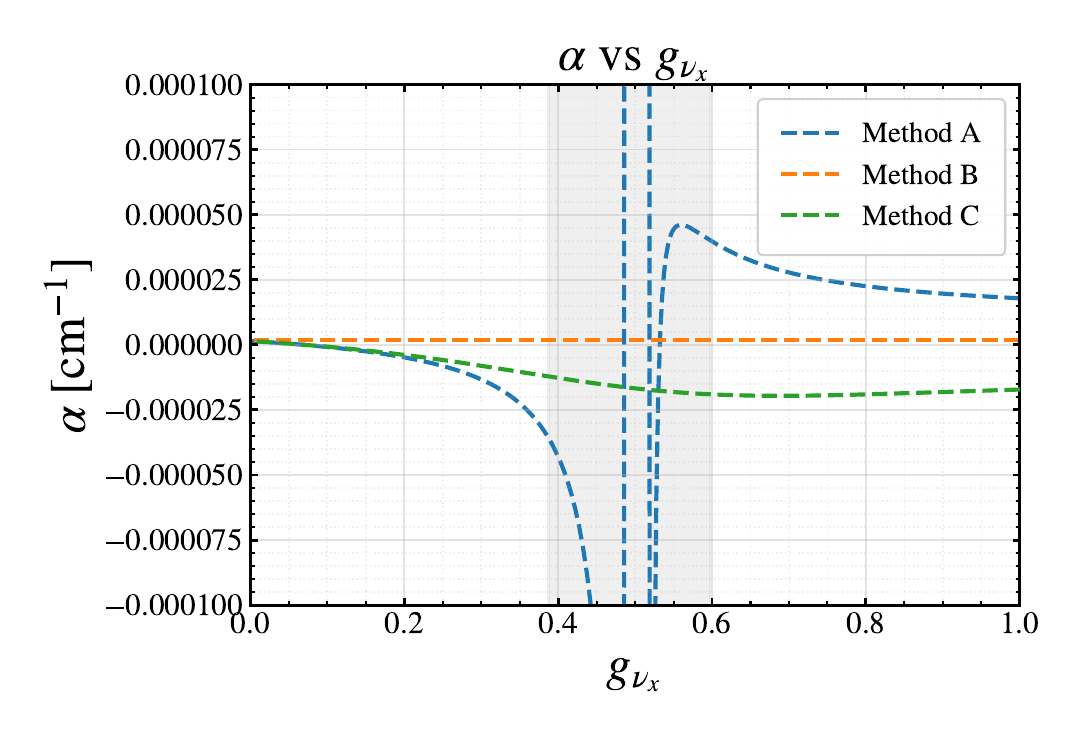}
    \caption{
    Effective collision rate parameters $\gamma^X$ (left) and $\alpha^X$ (right) for Model~I. The shaded region indicates strong cancellation in the energy integrals. Method~A exhibits divergences, while methods~B and~C remain finite.
    }
    \label{fig:model1_gammaalpha}
\end{figure*}

We first consider Model~I. Figure~\ref{fig:model1_gammaalpha} shows the effective collision rate parameters $\gamma^X$ and $\alpha^X$ as functions of $g_{\nu_x}$. The shaded region indicates strong cancellation in the energy integrals, quantified by
\begin{equation}
\begin{split}
    C_{\nu}\equiv\frac{\int E^2\,dE\,|\Delta f(E)|}{\left|\int E^2\,dE\,\Delta f(E)\right|},\qquad
    C_{\bar{\nu}}\equiv\frac{\int E^2\,dE\,|\Delta \bar{f}(E)|}{\left|\int E^2\,dE\,\Delta \bar{f}(E)\right|},
\end{split}
\end{equation}
with $\max(C_\nu,C_{\bar{\nu}})\ge3$.

In this regime, method~A develops divergences due to vanishing denominators in Eq.~\eqref{eq:Gamma_methodA}, leading to unphysical behavior in $\gamma^A$ and $\alpha^A$. Method~B remains finite but is insensitive to variations in $g_{\nu_x}$, since it depends only on the spectral shape. By contrast, method~C remains well behaved, producing smooth and finite effective parameters across the entire range.

\begin{figure*}[]
    \centering
    \includegraphics[width=0.48\linewidth]{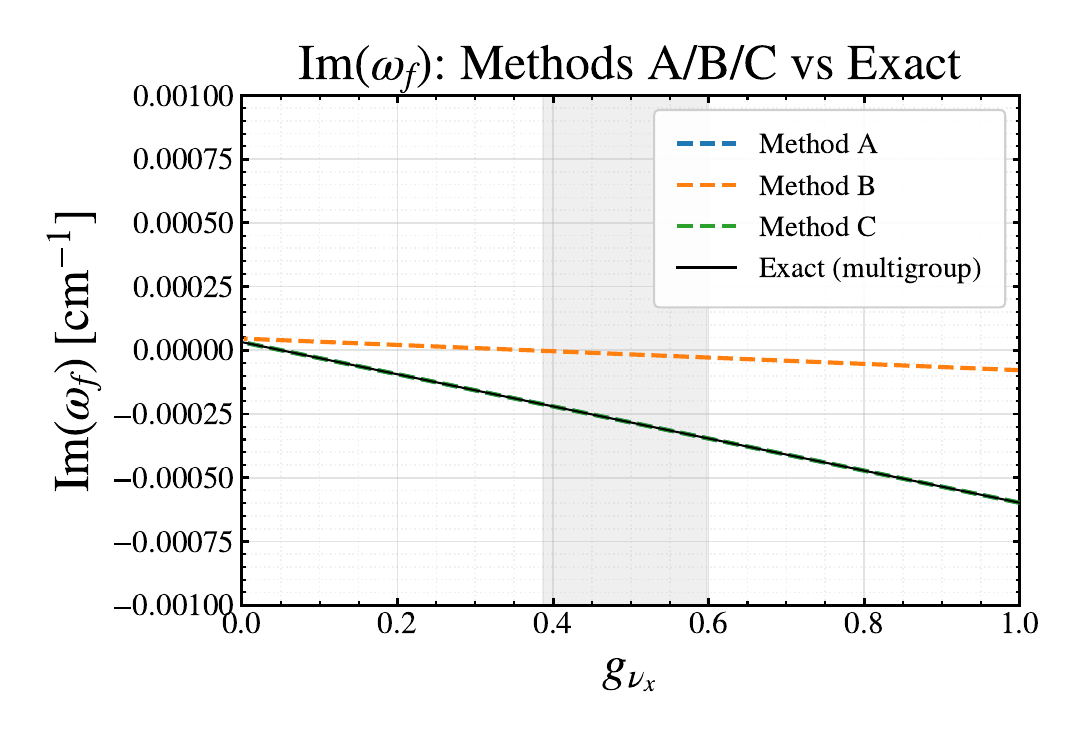}\hfill
    \includegraphics[width=0.48\linewidth]{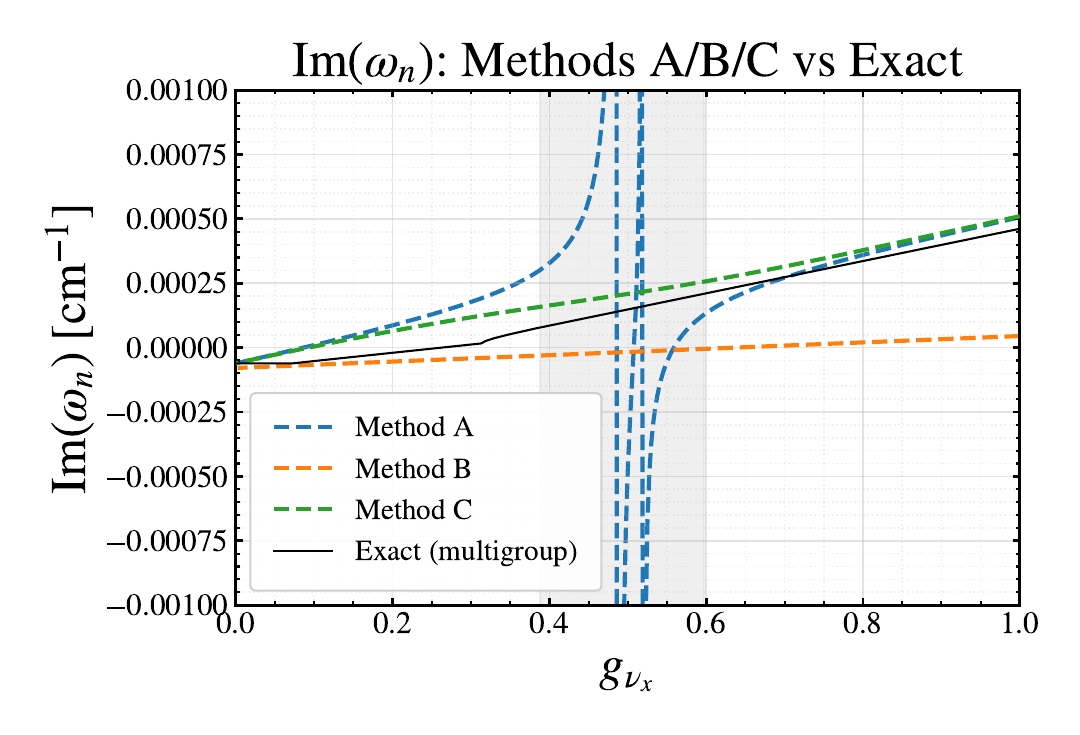}

    \vspace{0.5em}

    \includegraphics[width=0.48\linewidth]{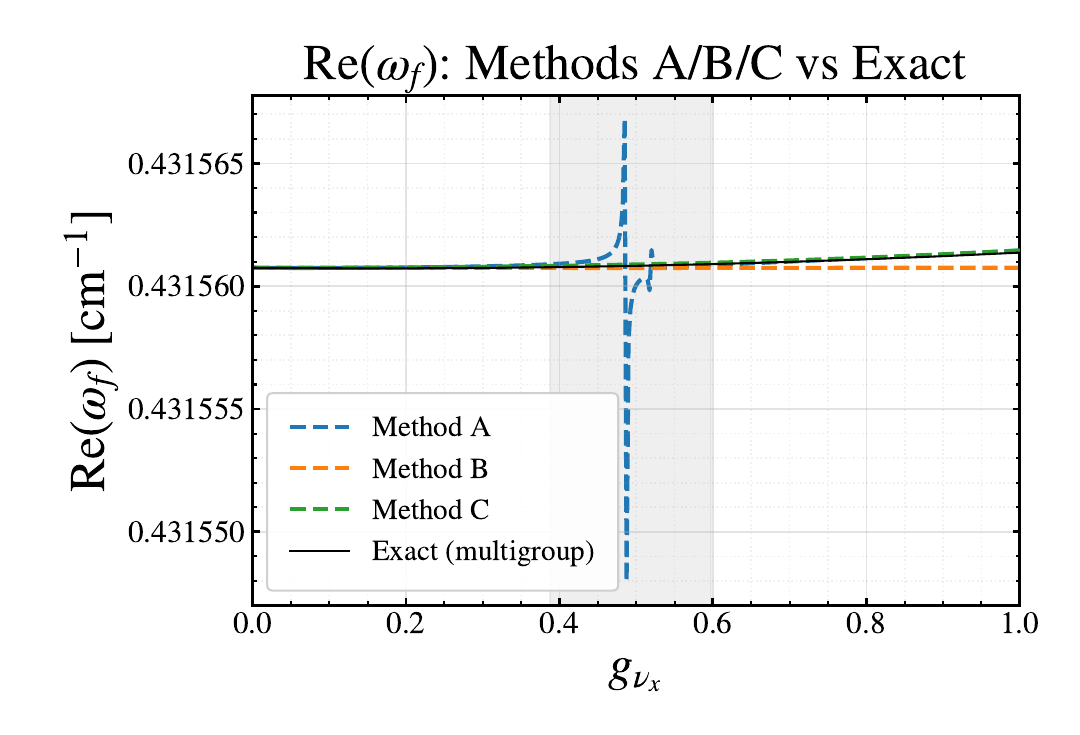}\hfill
    \includegraphics[width=0.48\linewidth]{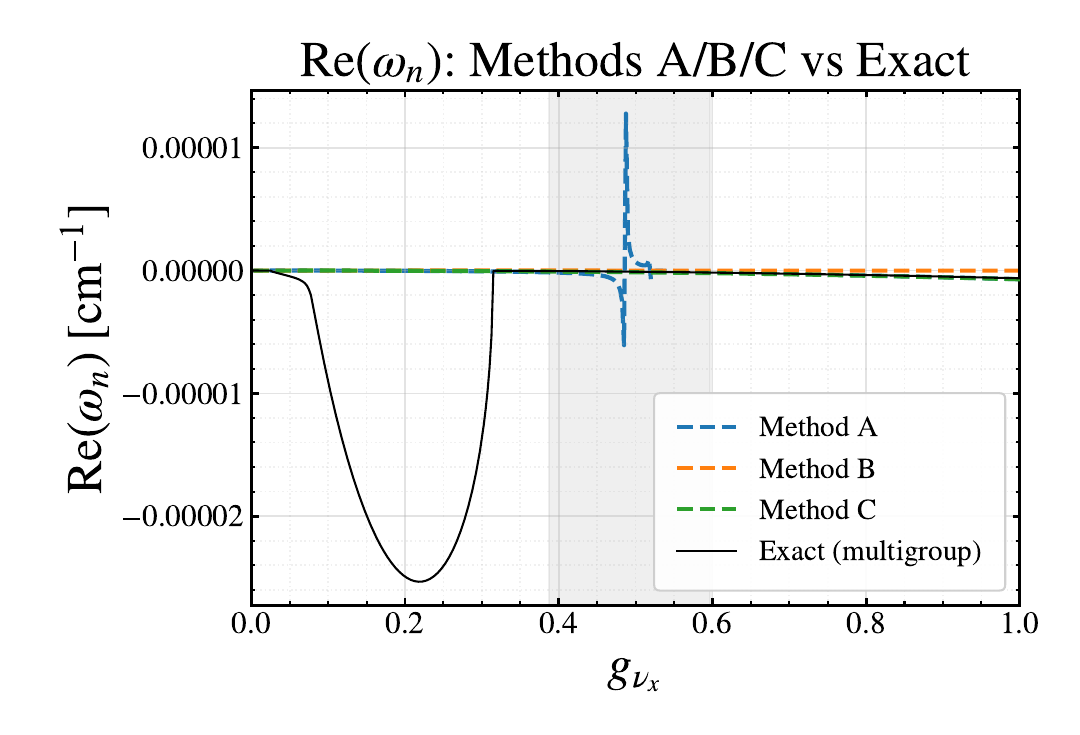}
    \caption{
    Real and imaginary parts of the collective eigenfrequencies for Model~I. Left and right panels correspond to far and near modes, respectively. In the upper left panel, the blue and green dashed lines closely overlap with the black solid curve. Method~A deviates strongly for the near mode in the strong-cancellation regime, while method~C remains accurate across the full range.
    }
    \label{fig:model1_roots}
\end{figure*}

Figure~\ref{fig:model1_roots} shows the real and imaginary parts of the two collective modes. As $g_{\nu_x}$ increases, the near mode becomes unstable while the far mode is suppressed. Method~A exhibits significant deviations in the near mode due to the divergence of its effective collision rates, while the far mode remains comparatively stable due to partial cancellation within the analytic solution. Method~B provides reasonable estimates at small $g_{\nu_x}$ but deviates at larger values. Method~C accurately reproduces both real and imaginary parts of the solutions across the full parameter range, with only minor discrepancies near the transition where the near mode changes stability.

We next consider Model~II, in which the parameters are tuned to satisfy $A^X=0$ (i.e., corresponding to resonance-like CFI). In this case, the two collective modes lie approximately symmetrically about the origin in the complex $\omega$ plane and are therefore labeled as mode~1 and mode~2.

\begin{figure*}[]
    \centering
    \includegraphics[width=0.48\linewidth]{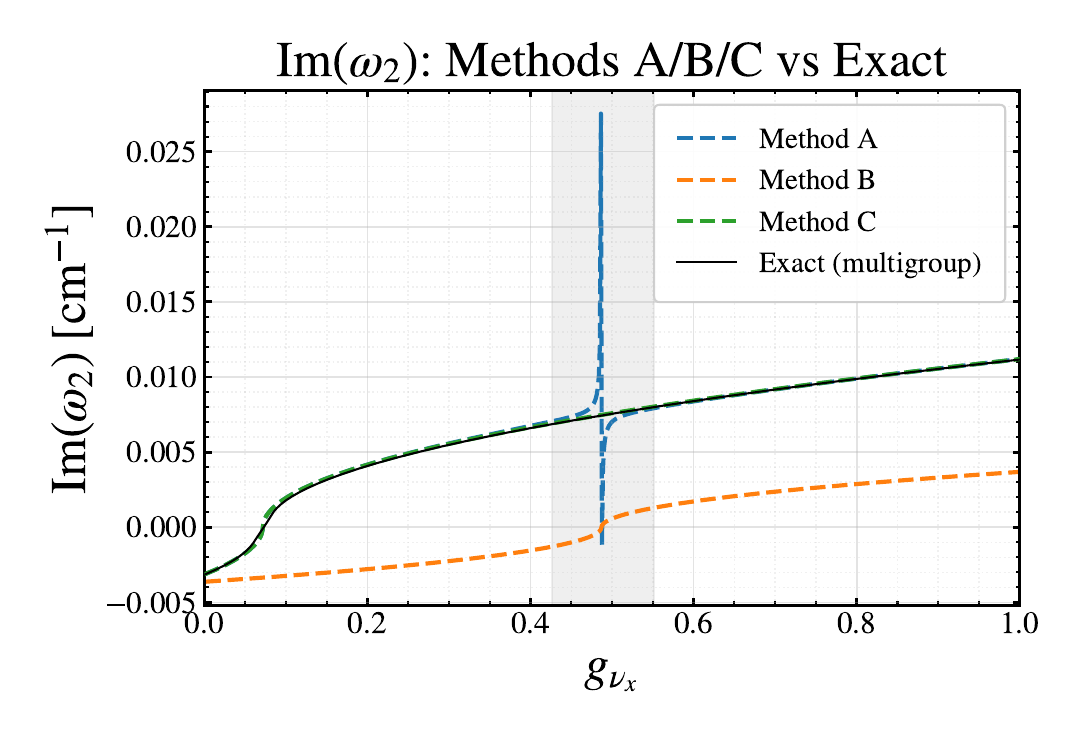}\hfill
    \includegraphics[width=0.48\linewidth]{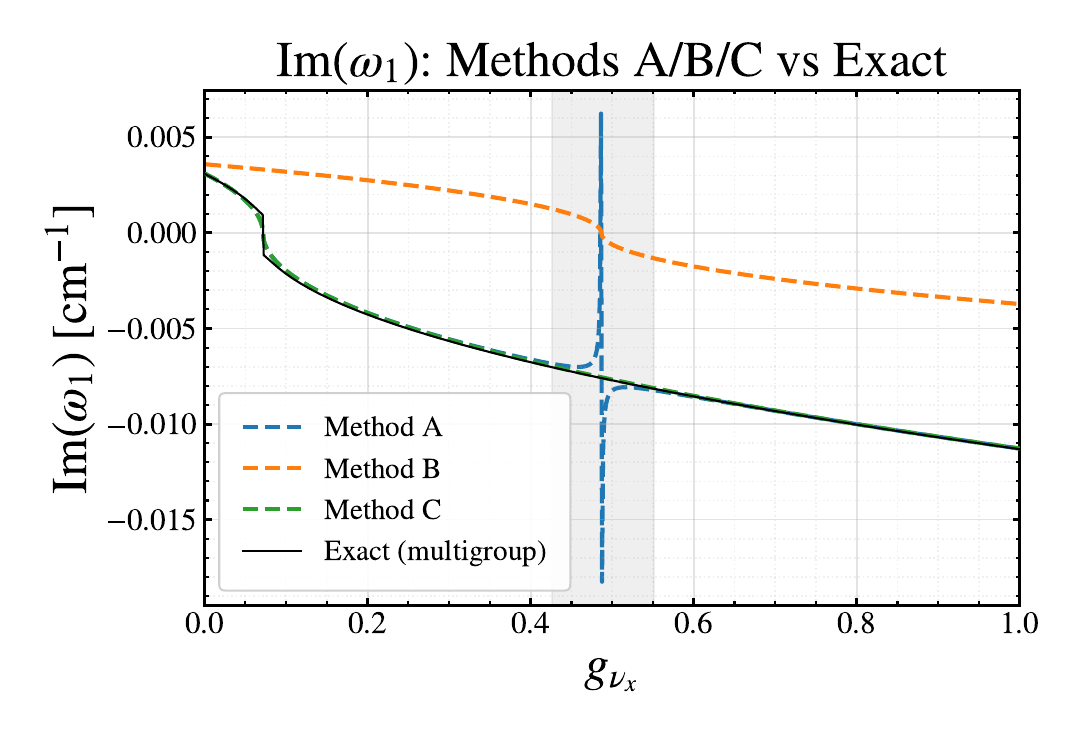}

    \vspace{0.5em}

    \includegraphics[width=0.48\linewidth]{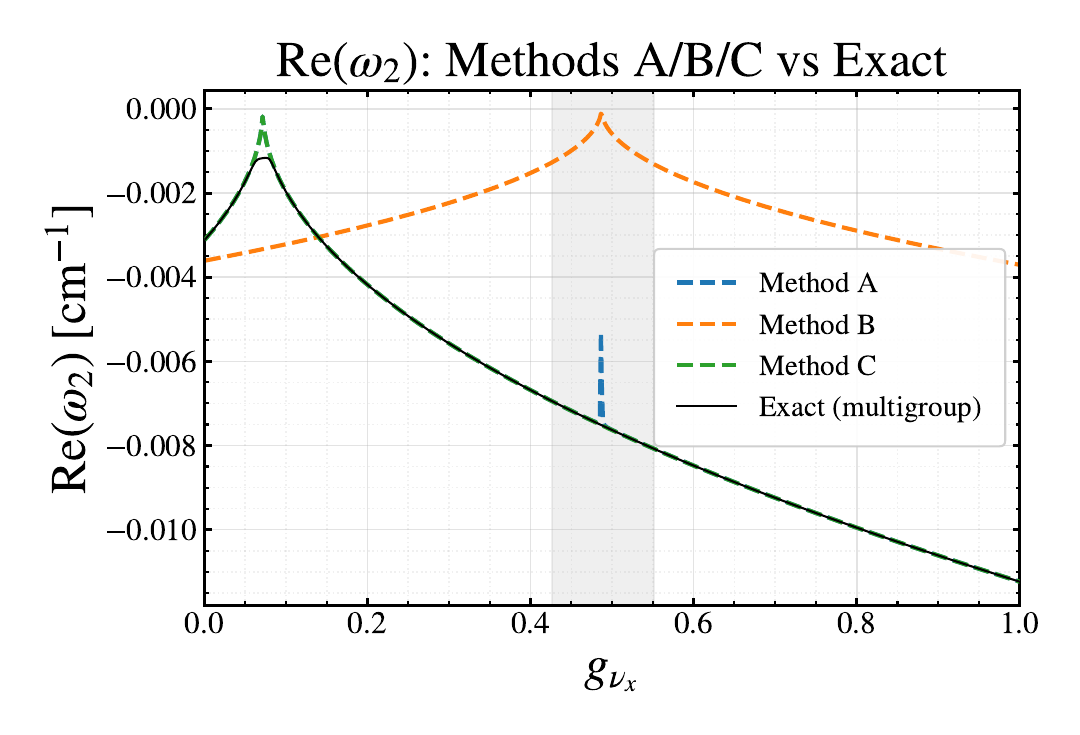}\hfill
    \includegraphics[width=0.48\linewidth]{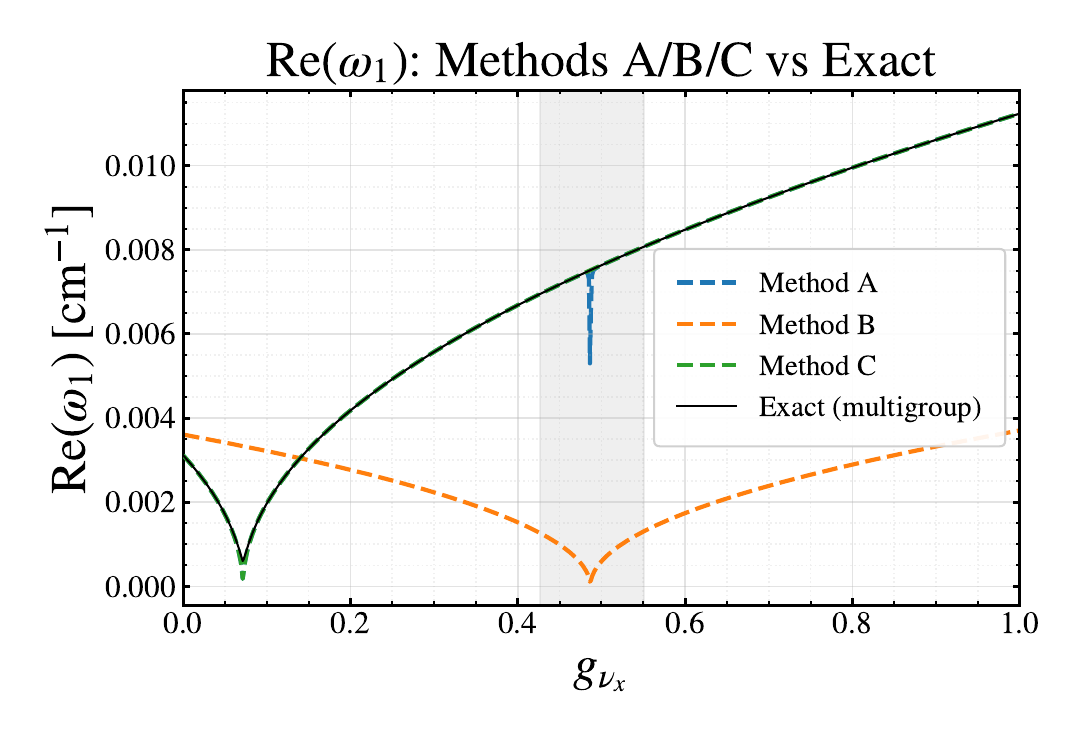}
    \caption{
    Same as Fig.~\ref{fig:model1_roots}, but for Model~II (resonance configuration). The two modes are labeled as mode~1 and mode~2.
    }
    \label{fig:model3_roots}
\end{figure*}

Figure~\ref{fig:model3_roots} shows the corresponding solutions. Method~A remains accurate except in regions where divergences occur, while method~B loses quantitative accuracy as the heavy-leptonic contribution ( $g_{\nu_x}$) increases. Method~C shows excellent agreement with the exact solution across the full parameter range.

Overall, method~C provides the most robust and accurate approximation across all models considered. Method~A can yield accurate predictions for the far mode in non-resonance regimes but suffers from divergences when strong cancellations occur. Method~B remains finite and qualitatively reasonable, but lacks a controlled analytical foundation and can deviate quantitatively at large $g_{\nu_x}$. 

By contrast, method~C consistently reproduces both real and imaginary parts of the collective modes across a wide range of spectral configurations, including cases with number cancellations. These results demonstrate that the sector-based reduction provides a reliable approximation to the full multi-energy dispersion relation.

\subsection{Anisotropic Distributions at $k=0$}
\label{subsec:Axi_k0}

In this subsection, we test the extension of the approximate energy-integration method to axisymmetric neutrino distributions. The dispersion-matrix components in Eqs.~\eqref{eq:Axi_Pi00}–\eqref{eq:Axi_Pi11} are evaluated using the reduced expressions in Eq.~\eqref{eq:Approx_Pi_Axi}. We restrict attention to $k=0$ modes and defer the extension to $k\neq0$ to the next subsection. The predictions of method~C are compared directly with exact solutions, while methods~A and~B are not shown, as their qualitative behavior follows that established in Sec.~\ref{subsec:Iso_k0}.

The Tr sector (see the definition around Eq.~\eqref{eq:TLTr_Pi} and the description below Eq.~\eqref{eq:Axi_Pi11}) satisfies $\Pi^{11}(\omega,0)=0$ and is formally identical to the IB branch of the isotropic problem under the mapping $\Delta f_0(E)/3-\Delta f_2(E)/15\mapsto\Delta f(E)$. The validity of method~C for the axisymmetric Tr sector therefore follows directly from its validity in the isotropic case. We therefore focus on the TL sector, which is governed by the condition $\Pi^{00}\Pi^{33}-(\Pi^{03})^2=0$.

We consider two regimes with axially symmetric angular distributions: (i) non-resonance CFIs, and (ii) resonance CFIs. The approximations of method~C are compared with exact solutions of the dispersion relation, with the angular structure treated analytically using Legendre moments.\footnote{One advantage of using Legendre moments is that modifying individual components does not affect CFIs in the isotropic limit, allowing a clean separation between isotropic collisional modes and anisotropic effects.}

Instead of specifying the full distributions $f_{\nu_\alpha}(E,v)$, we directly prescribe their first three Legendre moments. This is sufficient because only $\ell\in\{0,1,2\}$ enter the dispersion relation at $k=0$, as indicated by Eqs.~\eqref{eq:Axi_Pi00}-\eqref{eq:Axi_Pi11}. Analogous to Eq.~\eqref{eq:dist_iso}, we define
\begin{equation}
f_{\nu_\alpha,\ell}(E)=g_{\nu_\alpha,\ell}\,f_{\mathrm{FD}}(E;T_\alpha,\mu_\alpha),
\end{equation}
where $g_{\nu_\alpha,\ell}$ depends on species and Legendre index. While more general energy-dependent parametrizations are possible, this setup already captures the essential structure through crossings in $\Delta f_\ell$ and $\Delta \bar{f}_\ell$. The thermodynamic parameters $(T_\alpha,\mu_\alpha)$ are taken to be identical to those used in Sec.~\ref{subsec:Iso_k0} (see Table.~\ref{tab:param_iso}). The Legendre-moment scaling factors are summarized in Table~\ref{tab:param_axi}. In particular, the model NF\_NR does not exhibit any energy-integrated angular crossing that are necessary for fast flavor instability (FFI).

\begin{table}[]
\centering
\caption{Legendre-moment scaling parameters $g_{\nu_\alpha,\ell}$ used for the axisymmetric $k=0$ tests. Model names consist of two parts: NF denotes No FFI (Fast flavor instability) at $k=0$, and NR/R denotes Non-resonance/resonance CFIs. The constant $c_{\bar{\nu}_e}\simeq0.938$ is chosen to achieve the resonance condition $A=0$.}
\label{tab:param_axi}
\begin{tabular}{c c c c c c c c c c}
\hline\hline
Model
& $g_{\nu_e,0}$
& $g_{\bar{\nu}_e,0}$
& $g_{\nu_x,0}$
& $g_{\nu_e,1}$
& $g_{\bar{\nu}_e,1}$
& $g_{\nu_x,1}$
& $g_{\nu_e,2}$
& $g_{\bar{\nu}_e,2}$
& $g_{\nu_x,2}$\\
\hline
NF\_NR
& 1 & 1 & 0.5 & 0.1 & 0.15 & 0.1 & 0.2 & 0.25 & 0.15\\
\hline
NF\_R
& 1 & $c_{\bar{\nu}_e}$ & 0.5 & 0.13 & 0.12 & 0.1 & 0.35 & 0.2 & 0.15\\
\hline\hline
\end{tabular}
\end{table}

For each model, the dispersion relation is solved using the full energy dependence of the collision rates by adopting the same energy grid points used in the isotropic case (see Sec.~\ref{subsec:Iso_k0}). The angular structure is treated analytically via Legendre moments. We focus on the collective roots of the TL sector and compare their real and imaginary parts against the reduced solutions obtained using method~C.

As a reference, we also compare the axisymmetric modes with the corresponding Isotropized-CFI and pure FFI solution. Isotropized-CFIs are understood on isotropic angular distributions and obtained by isotropizing the distributions and evaluating the IP and IB branches using method~C. Pure FFIs are obtained in the collisionless limit $\Gamma(E)=\bar{\Gamma}(E)=0$, yielding analytic solutions in terms of the Legendre moments. Although the models presented in this subsection are designed to be fast stable, they play a role in influencing the CFI modes on anisotropic angular distributions.

\begin{figure*}[]
    \centering
    \includegraphics[width=\linewidth]{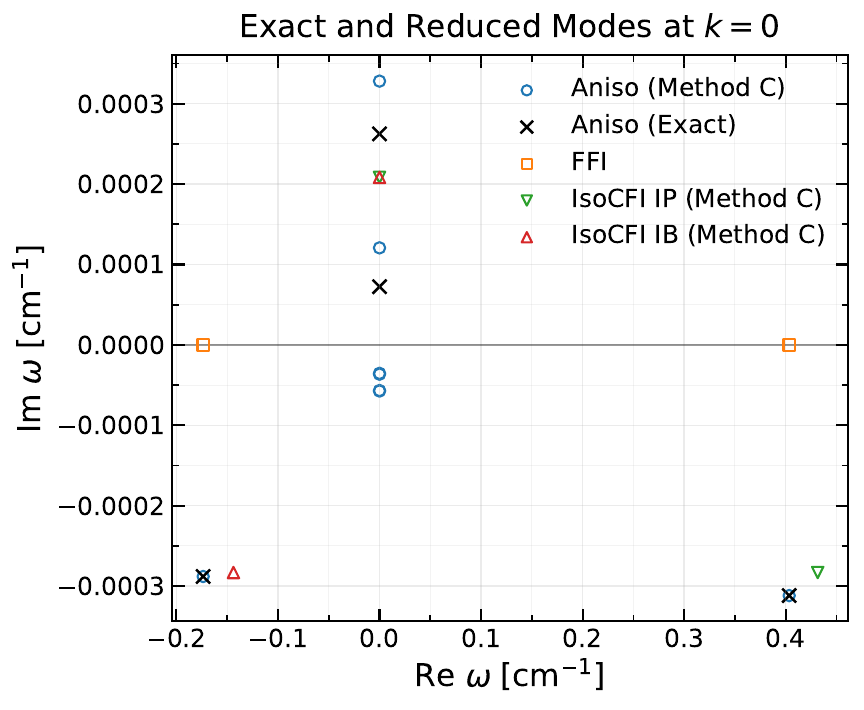}
    \caption{Exact and reduced axisymmetric modes at $k=0$ for Model~NF\_NR. Open blue circles show the solutions obtained using method~C, while black crosses denote the corresponding exact solutions. Open orange squares indicate pure fast modes. Open green downward triangles and open red upward triangles represent the Isotropized-CFI(IP) and Isotropized-CFI(IB) modes, respectively.
}
    \label{fig:Axi_NFNR}
\end{figure*}

We first consider Model~NF\_NR, shown in Fig.~\ref{fig:Axi_NFNR}. This model does not support unstable FFIs and exhibits non-resonance CFIs. Method~C accurately reproduces the two axisymmetric modes far from the origin, while the axisymmetric modes near the origin show small deviations due to their proximity to $\omega=0$. The anisotropic configuration causes the two axisymmetric near modes to be different from each other. The corresponding Isotropized-CFI modes lie close to the axisymmetric modes, but are nearly degenerate between the IP and IB branches.

\begin{figure*}[]
    \centering
    \includegraphics[width=\linewidth]{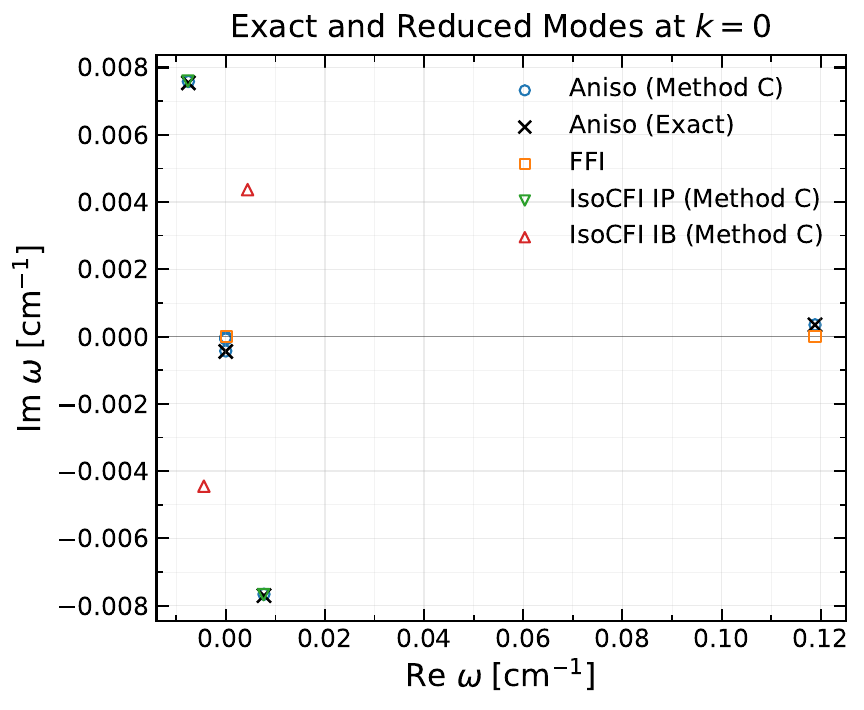}
    \caption{Same as Fig.~\ref{fig:Axi_NFNR}, but for Model~NF\_R.}
    \label{fig:Axi_NFR}
\end{figure*}

Next, we consider Model~NF\_R, shown in Fig.~\ref{fig:Axi_NFR}. This model lies in the resonance CFI regime but still does not support FFIs. Method~C reproduces the axisymmetric modes with high accuracy. In this case, the two mode pairs respond differently to the angular structure, with one pair remaining close to their isotropic counterparts while the other exhibits stronger sensitivity and partial overlap with the fast modes.

In addition to the two models without FFIs at $k=0$, we provide two models with both FFIs and CFIs at $k=0$ in Appendix~\ref{app:FFI_CFI_k0} to further demonstrate the robustness of method~C in effectively averaging over the energy degree of freedom. Across all models considered in this paper, method~C reproduces the exact axisymmetric CFI modes with good quantitative accuracy. The largest deviations occur for modes located very close to the origin. As in the isotropic case, these residual errors arise from the proximity of the modes to the origin, where the approximation becomes less accurate; nevertheless, the resulting accuracy remains reasonable for practical purposes.

These results demonstrate that method~C provides a robust and computationally efficient framework for analyzing collective flavor instabilities in axisymmetric systems at $k=0$. Its ability to accurately capture the instability landscape without resolving the full energy dependence makes it particularly well suited for large-scale parameter surveys and exploratory studies.

\subsection{Extension to Inhomogeneous Modes ($k \neq 0$)}
\label{subsec:knon0}

We now extend our analysis to inhomogeneous modes with $k \neq 0$. In this regime, the streaming term $kv$ couples the angular and frequency dependences in the dispersion relation, rendering the effective collision rates intrinsically angle dependent. Consequently, the separation between collisional and fast modes, which is transparent at $k=0$, is no longer explicit, and the validity of reduced energy-integration schemes is not guaranteed \emph{a priori}. In this subsection, we demonstrate that method~C [Eq.~\eqref{eq:effGm_knon0}], applied separately to the positive- and negative-energy sectors, continues to provide accurate predictions for collective modes despite this additional coupling.

\begin{figure*}[]
\centering
\includegraphics[width=0.49\linewidth]{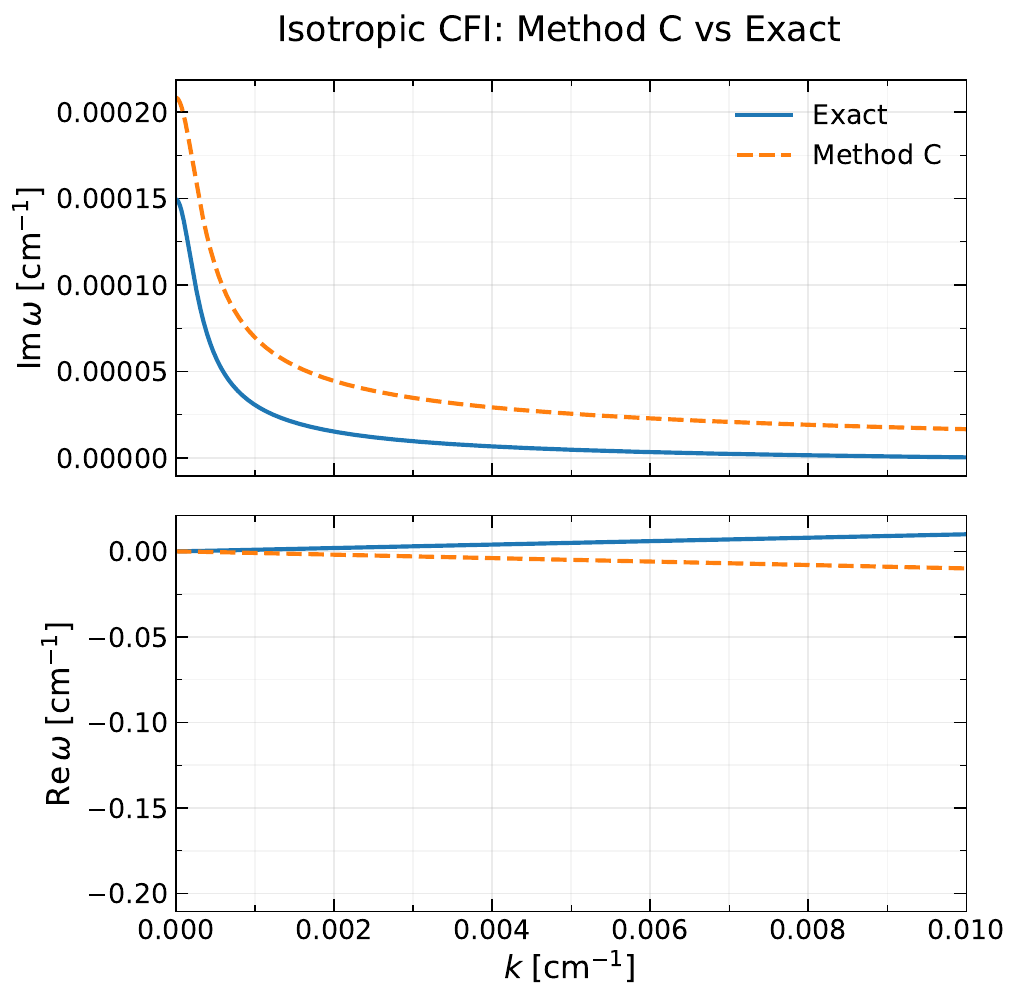}
\includegraphics[width=0.49\linewidth]{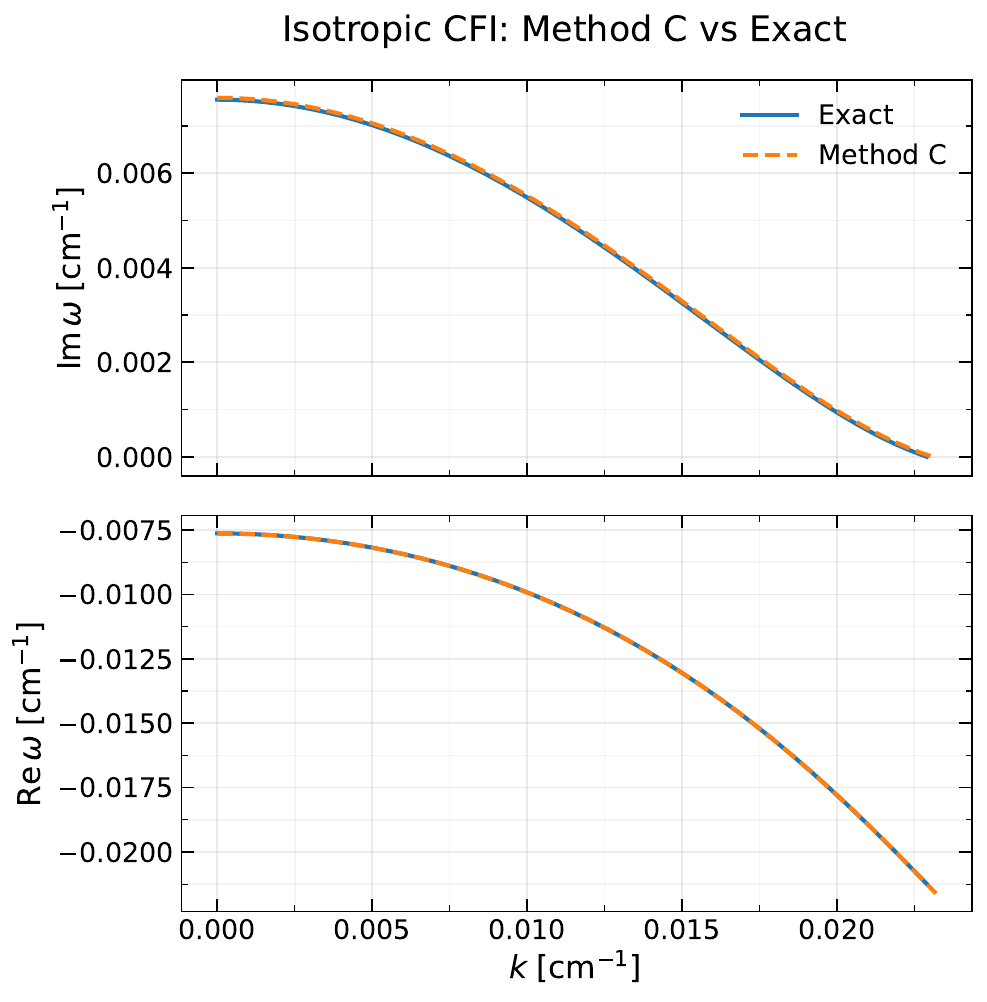}
\caption{
Comparison of dispersion relations $\omega(k)$ obtained with method~C and from exact calculations for isotropic models (see main text for model details). Left panels: non-resonance CFI model (near mode). Right panels: resonance CFI model. Top: imaginary part of $\omega(k)$ (growth rate). Bottom: real part of $\omega(k)$. Method~C (orange dashed) closely reproduces the exact solutions (blue solid) for both branches over the full $k$ range.
}
\label{fig:knon0_IsoCFI}
\end{figure*}

We consider representative and nontrivial cases and compare the method~C approximation against exact solutions for both isotropic and anisotropic angular distributions. We begin with isotropic models. For these tests, we adopt the two models introduced in Sec.~\ref{subsec:Iso_k0}, fixing the parameter $g_{\nu_x}=0.5$ (see Eqs.~\eqref{eq:dist_iso}–\eqref{eq:Gamma_params} and Table~\ref{tab:param_iso} for the model setup). The comparison between method~C and the exact solutions is shown in Fig.~\ref{fig:knon0_IsoCFI}.

For the non-resonance CFI model (left panels), the solution corresponds to the near mode. Although a few tens of percent deviations are involved, method~C qualitatively captures the trend in the exact solution. For the resonance CFI model (right panels), the agreement is excellent across the full $k$ range. If the non-resonance CFI mode were a far mode, the approximation would also agree excellently with the exact solution across the full $k$ range.

\begin{figure*}[]
\centering
\includegraphics[width=0.49\linewidth]{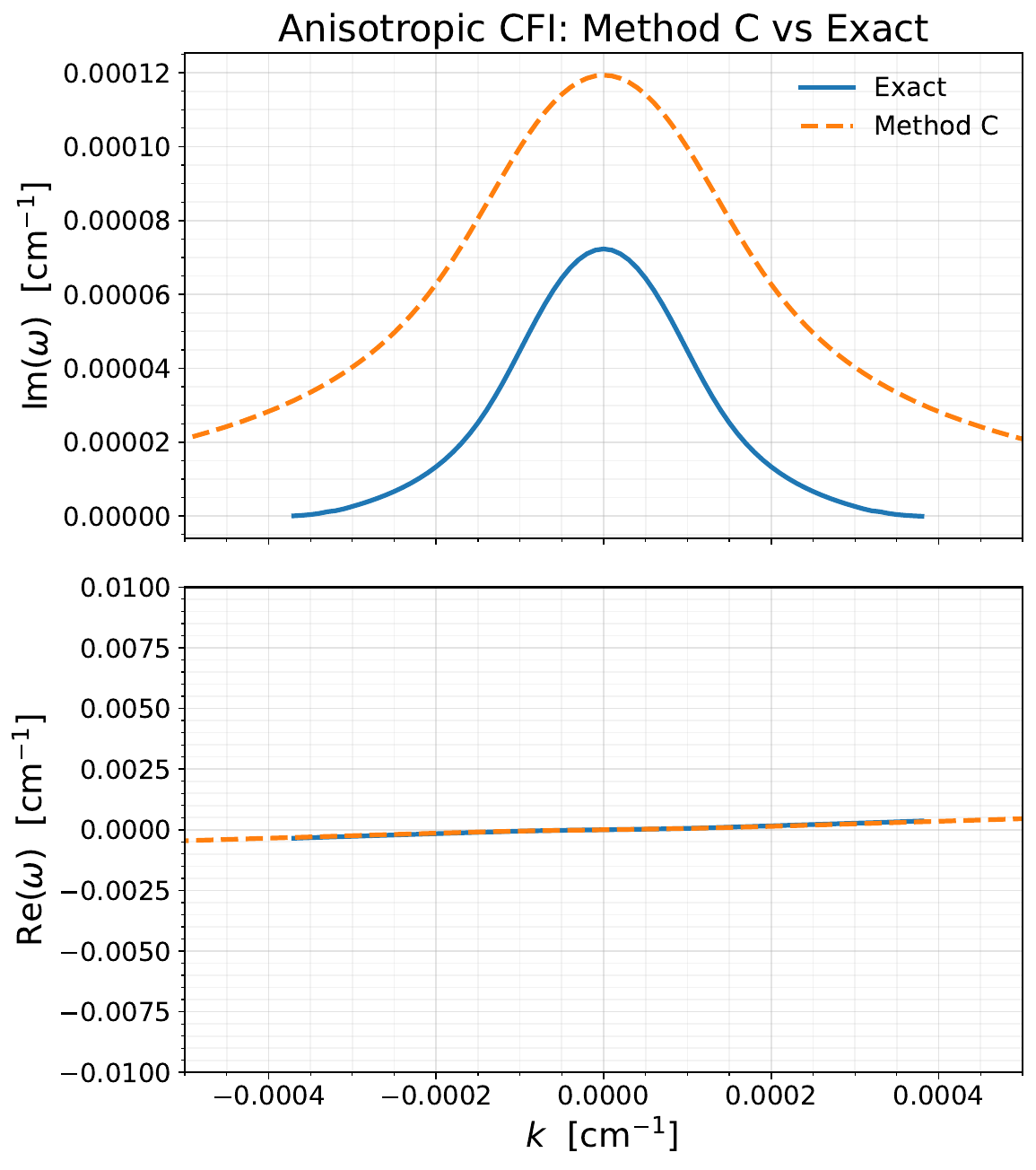}
\includegraphics[width=0.49\linewidth]{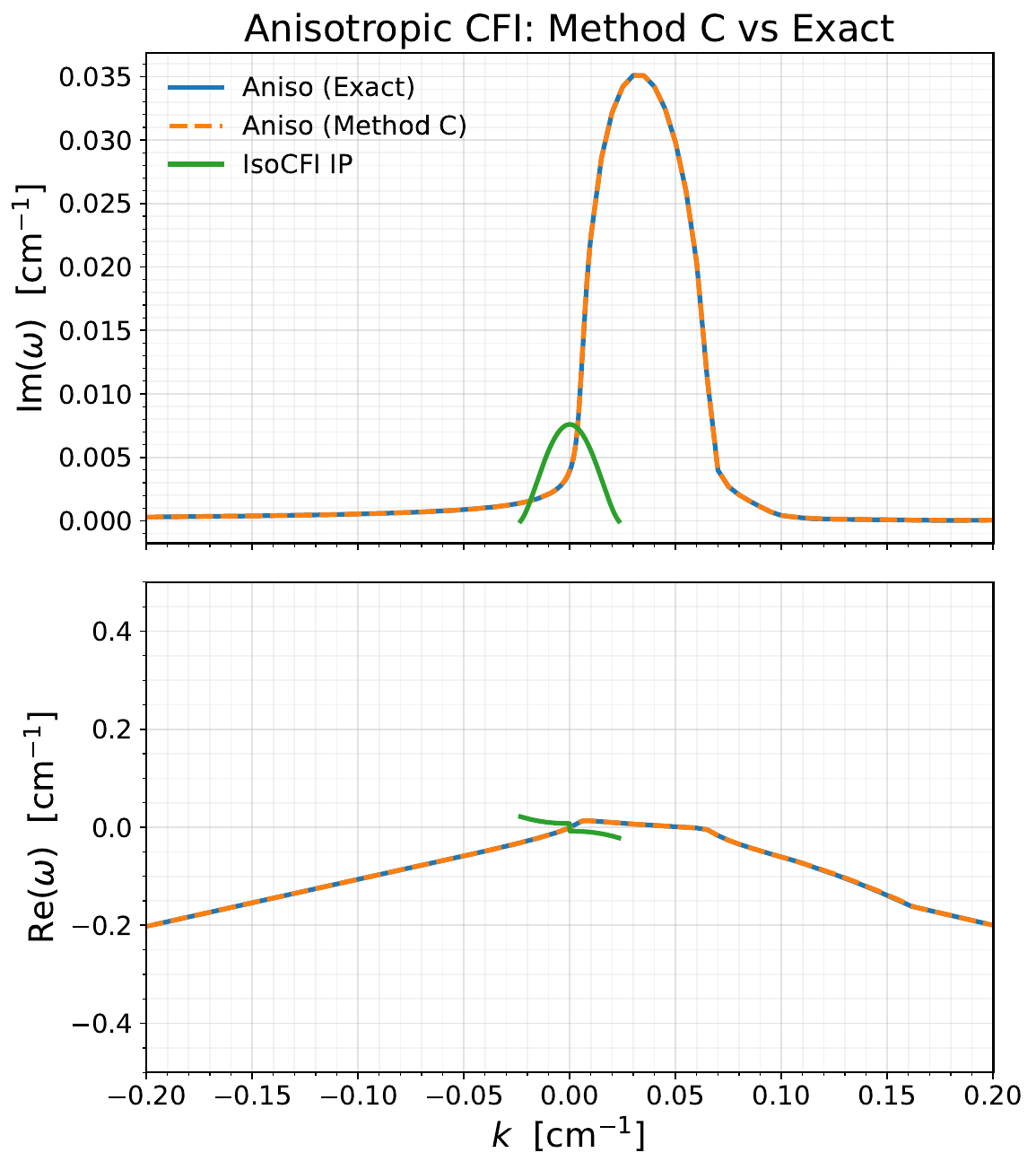}
\caption{
Comparison of dispersion relations $\omega(k)$ obtained with method~C and from exact calculations for axisymmetric models (see main text for model details). Left panels: model NF\_NR. Right panels: model F\_R. Top: imaginary part of $\omega(k)$ (growth rate). Bottom: real part of $\omega(k)$. Method~C (orange dashed) closely reproduces the exact solutions (blue solid) for the collisional branches over the full $k$ range. In the right panels, the anisotropic collisional mode is additionally compared with its Isotropized-CFI counterpart (green solid).
}
\label{fig:knon0_AxiCFI}
\end{figure*}

We next consider the case with anisotropic angular distributions. We adopt the unstable collisional modes from the axisymmetric models NF\_NR and F\_R (see Tables~\ref{tab:param_axi} and \ref{tab:param_axi2} for model parameters; the model NF\_NR is given in Sec.~\ref{subsec:Axi_k0} and the model F\_R is appended in the appendix~\ref{app:FFI_CFI_k0}). The comparison between method~C and the exact solutions is shown in Fig.~\ref{fig:knon0_AxiCFI}.
For model NF\_NR (left panels), no angular crossings are present, and all branches correspond to collisional modes. We focus on the less unstable branch (see Fig.~\ref{fig:Axi_NFNR}). A roughly constant deviation persists over the entire $k$ range, reflecting the near-mode nature of this CFI branch. Nevertheless, the approximation remains sufficiently accurate for practical purposes.
For model F\_R (right panels), the near unstable mode corresponds to a collisional mode across the full $k$ range (see the right panel of Fig.~\ref{fig:Axi_FNR_FR}). At $k=0$, this model exhibits two unstable collective modes, but only the near mode is collisional. We therefore compare the method~C prediction for this mode with the exact solution in Fig.~\ref{fig:knon0_AxiCFI}. The agreement is excellent.

We further compare the axisymmetric CFI with its counterpart in the isotropic limit, corresponding to the Isotropized-CFI IP mode.\footnote{In the collisionless limit, this mode becomes stable and is therefore classified as a collisional mode rather than a fast mode.} Axisymmetry significantly enhances the maximum growth rate and shifts the location of the peak in $k$. Previous surveys of CFIs in CCSN and BNSM models may have underestimated the growth rates by restricting attention to Isotropized-CFIs. The mechanism underlying this enhancement of collisional modes warrants further investigation and will be presented in a subsequent paper.

These results demonstrate that method~C remains robust beyond the $k=0$ limit, even in regimes where axisymmetry-induced effects are strong. Its ability to accurately capture collisional modes across a range of coupling regimes makes it a practical and reliable tool for dispersion-relation analysis in systems exhibiting CFI.

\section{Conclusions}
\label{sec:Conclusion}

In this work, we developed a new approximate energy-integration method for analyzing CFI in dense neutrino media. The method is designed to reduce the multi-energy dispersion relation to a compact form while retaining the essential spectral structure that controls the instability. In contrast to previous reduced treatments, the construction is based on a sector-wise decomposition of the signed neutrino and antineutrino contributions, which avoids cancellations within individual effective sectors. As a result, the effective collision rates remain nonnegative and well defined even in the presence of energy crossings, eliminating the spurious singular behavior that can arise in earlier schemes.

We first formulated the method in the simplest setting of isotropic angular distributions and homogeneous modes. In this limit, the reduced dispersion relation preserves the algebraic simplicity of previous approximations while providing a more controlled description of the underlying multi-energy problem. We then extended the construction to axisymmetric backgrounds and to inhomogeneous modes with $k\neq0$, where the coupling between angular structure and frequency makes the reduction problem substantially less trivial. Although the treatment of angular structure was not reduced as aggressively as the energy dependence, the same sector-based strategy continues to provide a practical approximation framework beyond the isotropic $k=0$ limit.

Through a series of controlled numerical experiments, we assessed the performance of the method across isotropic and anisotropic models, including both homogeneous and inhomogeneous cases. In the isotropic $k=0$ limit, method~C consistently outperforms the two previous approximate schemes: method~A can become ill defined or quantitatively unreliable in the presence of strong spectral cancellations, while method~B, although finite, lacks a controlled derivation and can deviate significantly in realistic regimes. By contrast, method~C reproduces both the real frequencies and growth rates of unstable collisional modes with good quantitative accuracy across a broad range of models. The same conclusion continues to hold for axisymmetric backgrounds and for modes with $k\neq0$, where the method remains robust even when axisymmetry-induced effects substantially modify the instability structure.

Our analysis also clarified the main regime in which the approximation becomes less accurate. The largest deviations occur for modes located close to the origin in the complex $\omega$-plane, namely the near modes in non-resonance configurations. This behavior follows naturally from the structure of the expansion, whose accuracy is controlled by the size of the effective denominators relative to the collision rates. Modes with larger characteristic frequency scale, including far modes and resonance modes sufficiently separated from the origin, are generally reproduced with much higher accuracy. This geometric interpretation provides a useful diagnostic for assessing the expected reliability of the approximation in practical applications.

The present work establishes that the energy dependence of CFI can be reduced in a controlled and robust manner without sacrificing the key physical structure of the problem. This makes method~C a useful tool for systematic surveys of CFI in realistic neutrino distributions, where repeated solution of the full multi-energy and multi-angle dispersion relation would otherwise be operationally cumbersome. 

Several extensions remain for future work. In particular, it would be desirable to develop a comparably effective reduction of the angular structure for general inhomogeneous modes, to understand more fully the enhancement of collisional modes in axisymmetric systems, and to apply the method to neutrino distributions extracted from state-of-the-art CCSN and BNSM simulations. These issues are currently under investigation and will be addressed elsewhere.

\acknowledgments
We are grateful to Masamichi Zaizen, Lucas Johns, and Tianshu Wang for useful comments and discussions. H.N. is supported by Grant-in-Aid for Scientific Research (23K03468), the NINS International Research Exchange Support Program, and the HPCI System Research Project (Project ID: hp250006, hp250226, hp250166, hp260058).

\bibliographystyle{apsrev4-2}
\bibliography{ref}


\appendix
\section{Coexistence of Fast and Collisional Modes at $k=0$}
\label{app:FFI_CFI_k0}
In this appendix, we present two additional axisymmetric models that exhibit both FFIs and CFIs at $k=0$. The models are constructed in the same manner as those in Sec.~\ref{subsec:Axi_k0}, and their parameters are summarized in Table~\ref{tab:param_axi2}.

\begin{figure*}[]
    \centering
    \includegraphics[width=0.49\linewidth]{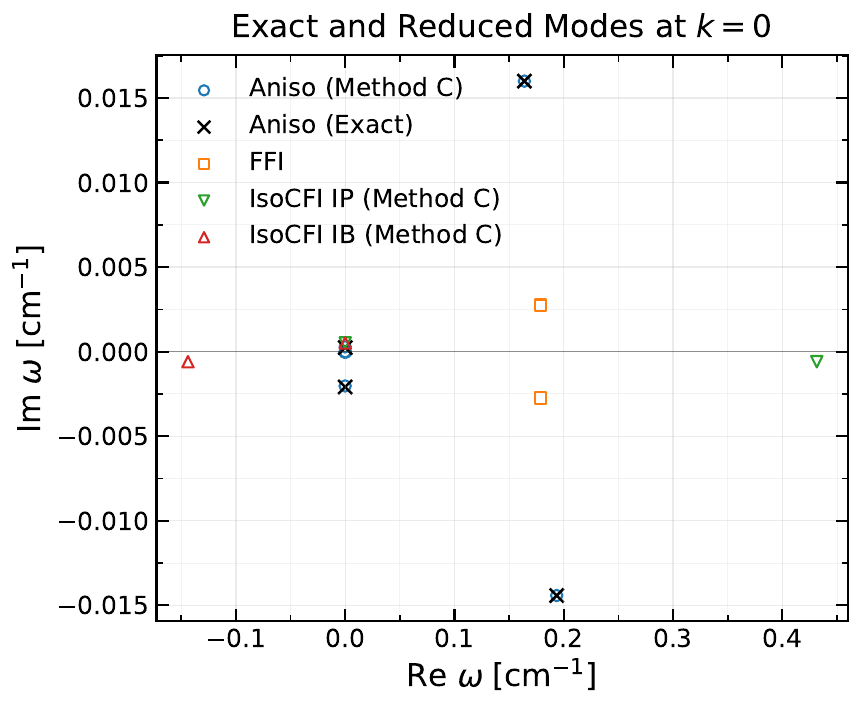}
    \includegraphics[width=0.49\linewidth]{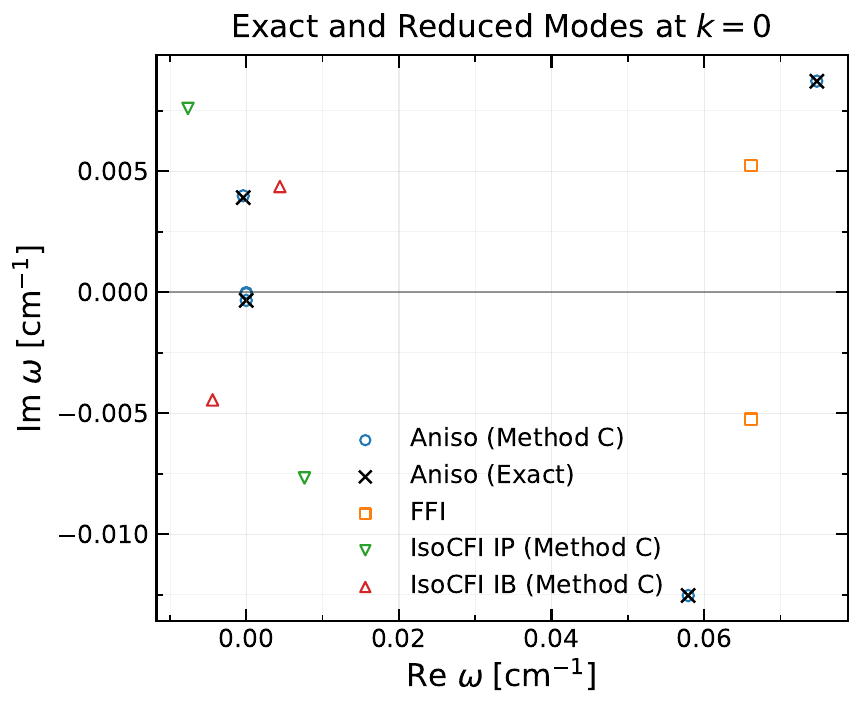}
    \caption{Same as Figs.~\ref{fig:Axi_NFNR} and \ref{fig:Axi_NFR}, but for models F\_NR (left panel) and F\_R (right panel). The model parameters are given in Appendix~\ref{app:FFI_CFI_k0}. 
}
    \label{fig:Axi_FNR_FR}
\end{figure*}

The approximate and exact solutions are shown in Fig.~\ref{fig:Axi_FNR_FR}. Despite the presence of FFIs and the resulting increased complexity, method~C maintains excellent agreement with the exact solutions for all axisymmetric modes. 

\begin{table}[]
\centering
\caption{Legendre-moment scaling parameters $g_{\nu_\alpha,\ell}$ used for the axisymmetric $k=0$ tests. Model names consist of two parts: F denotes models in which FFIs are unstable at $k=0$, and NR/R denotes non-resonance/resonance CFIs. The constant $c_{\bar{\nu}_e}\simeq0.938$ is chosen to satisfy the resonance condition $A=0$.}
\label{tab:param_axi2}
\begin{tabular}{c c c c c c c c c c}
\hline\hline
Model
& $g_{\nu_e,0}$
& $g_{\bar{\nu}_e,0}$
& $g_{\nu_x,0}$
& $g_{\nu_e,1}$
& $g_{\bar{\nu}_e,1}$
& $g_{\nu_x,1}$
& $g_{\nu_e,2}$
& $g_{\bar{\nu}_e,2}$
& $g_{\nu_x,2}$\\
\hline
F\_NR
& 1 & 1 & 0.5 & 0.3 & 0.39041 & 0.15 & 0.4 & 0.3 & 0.15\\
\hline
F\_R
& 1 & $c_{\bar{\nu}_e}$ & 0.5 & 0.3 & 0.31 & 0.15 & 0.45 & 0.2795 & 0.15\\
\hline\hline
\end{tabular}
\end{table}

\end{document}